\documentclass[fleqn,usenatbib]{mnras}

\usepackage{newtxtext,newtxmath,color}
\usepackage{amsmath, graphics, setspace, bm, graphicx}
\allowdisplaybreaks

\defcitealias{DawsonChiang(2014)}{DC14}

\newcommand{\mathsym}[1]{{}}
\newcommand{\unicode}[1]{{}}

\newcommand{\Sg}{\Sigma}

\newcommand{\dg}{\delta}
\newcommand{\Dg}{\Delta}

\newcommand{\Om}{\Omega}
\newcommand{\om}{\omega}
\newcommand{\pom}{\varpi}

\newcommand{\der}{{\rm d}}

\newcommand{\mdo}{m_{\rm d0}}
\newcommand{\ado}{a_{\rm d0}}

\newcommand{\e}{{\rm e}}

\newcommand{\au}{{\rm au}}

\newcommand{\rin}{r_{\rm in}}
\newcommand{\rout}{r_{\rm out}}
\newcommand{\ve}{{\bm e}}

\newcommand{\hl}{{\bm {\hat l}}}
\newcommand{\hld}{{\bm {\hat l}}_{\rm d}}

\newcommand{\hs}{{\bm {\hat s}}}

\newcommand{\Myr}{{\rm Myr}}

\newcommand{\ms}{M_\star}

\newcommand{\md}{m_{\rm d}}

\newcommand{\Mjup}{{\rm m}_{\rm J}}

\newcommand{\roin}{r_{\rm 1,in}}
\newcommand{\roout}{r_{\rm 1,out}}
\newcommand{\rtin}{r_{\rm 2,in}}
\newcommand{\rtout}{r_{\rm 2,out}}

\newcommand{\Msun}{{\rm M}_\odot}
\newcommand{\Rsun}{{\rm R}_\odot}

\newcommand{\btimes}{{\bm \times}}
\newcommand{\bcdot}{{\bm \cdot}}

\newcommand{\yr}{{\rm yr}}

\newcommand{\apl}{a_{\rm p}}

\newcommand{\hlpl}{{\bm {\hat l}}_{\rm p}}

\newcommand{\be}{\begin{equation}}
\newcommand{\ee}{\end{equation}}

\usepackage[T1]{fontenc}

\DeclareRobustCommand{\VAN}[3]{#2}
\let\VANthebibliography\thebibliography
\def\thebibliography{\DeclareRobustCommand{\VAN}[3]{##3}\VANthebibliography}

\usepackage{graphicx}	
\usepackage{amsmath}	

\title[Secular resonances in transition discs]{Sweeping Secular Resonances and Giant Planet Inclinations \\in Transition Discs}

\author[Zanazzi \& Chiang]{J. J. Zanazzi$^{1}$\thanks{51 Pegasi b fellow, email:jzanazzi@berkeley.edu} and E.~Chiang$^{1,2}$ \\
$^{1}$Astronomy Department, Theoretical Astrophysics Center, and Center for Integrative Planetary Science, University of California
Berkeley, \\
Berkeley, CA 94720, USA \\
$^{2}$Department of Earth and Planetary Science, University of California, Berkeley, CA 94720, USA}

\date{Accepted XXX. Received YYY; in original form ZZZ}

\pubyear{2023}

\begin{document}
\maketitle


\begin{abstract}
    The orbits of some warm Jupiters are highly inclined (20$^\circ$--50$^\circ$) to those of their exterior companions. Comparable misalignments are inferred between the outer and inner portions of some transition discs. These large inclinations may originate from planet-planet and planet-disc secular resonances that sweep across interplanetary space as parent discs disperse. The maximum factor by which a seed mutual inclination can be amplified is of order the square root of the angular momentum ratio of the resonant pair. We identify those giant planet systems (e.g.~Kepler-448 and Kepler-693) which may have crossed a secular resonance, and estimate the required planet masses and semimajor axes in transition discs needed to warp their innermost portions (e.g.~in CQ Tau). Passage through an inclination secular resonance could also explain the hypothesized large mutual inclinations in apsidally-orthogonal warm Jupiter systems (e.g.~HD 147018).
\end{abstract}

\section{Introduction}

Most planetary systems are flat \citep[e.g.][]{WinnFabrycky(2015),ZhuDong(2021)}, but some are not. In addition to large stellar obliquities measured for single planets using the Rossiter-McLaughlin effect (e.g. \citealt{Albrecht+(2022),Siegel+(2023),DongForeman-Mackey(2023)}), large mutual inclinations between planets have been suspected or confirmed. \cite{DawsonChiang(2014)} proposed that certain warm Jupiters are inclined by $i_{\rm mut} \approx 40^\circ$ relative to exterior super-Jupiter companions, based on their relative apsidal orientations. Transit duration variations imply substantial misalignments in the warm Jupiter systems 
Kepler-448 ($i_{\rm mut} = {20_{-12}^{+17}}^\circ$, \citealt{Masuda(2017)}), Kepler-693 ($i_{\rm mut} = {53_{-9}^{+7}}^\circ$, \citealt{Masuda(2017)}), Kepler-108 ($i_{\rm mut} = {24_{-8}^{+11}}^\circ$, \citealt{MillsFabrycky(2017)}), and WASP-148 ($i_{\rm mut} = {21_{-5}^{+5}}^\circ$, \citealt{Almenara+(2022)}). The inclination between a cold Jupiter and inner super-Earth in $\pi$ Men is astrometrically constrained to lie between $49^\circ < i_{\rm mut} < 131^\circ$ \citep{XuanWyatt(2020)}.

Some protoplanetary discs also exhibit large misalignments/warps. More than a dozen transitional discs (defined as having large cavities) are observed in scattered light to be shadowed by circumstellar material closer to their host stars (\citealt{Benisty+(2022)}). Azimuthally extended shadows are cast by close-in discs inclined by $\sim$$10^\circ$ relative to their outer discs (e.g.~\citealt{Stolker+(2016),Stolker+(2017),Debes+(2017),Muro-Arena+(2020)}), while narrow, diametrically opposed shadows are cast by the nodes of more highly inclined inner discs ($\sim$$30^\circ-90^\circ$; e.g.~\citealt{Marino+(2015),Benisty+(2017),Long+(2017),Casassus+(2018),Pinilla+(2018),Uyama+(2020),Ginski+(2021)}). Supporting evidence for misaligned discs comes 
\newpage\noindent from interferometric imaging \citep[e.g.][]{Kluska+(2020),GRAVITY(2021a),Bohn+(2022)} and CO kinematics \citep[e.g.][]{Casassus+(2015),Loomis+(2017),Mayama+(2018),Perez+(2018),Bi+(2020),Kraus+(2020)}.

One mechanism for exciting inclinations is secular resonance. When two bodies precess nodally at the same rate, angular momentum can be efficiently transferred between them, lifting one orbit while lowering the other and amplifying $i_{\rm mut}$ in the net. Analogously, when apsidal precession rates match,  eccentricities can change dramatically. Two planets can be driven through a secular resonance by their parent protoplanetary disc; as disc material depletes, planetary precession frequencies change, and can momentarily match. Secular resonance crossings driven by the depletion of the solar nebula may have excited the inclinations and eccentricities of the terrestrial planets, asteroids, and Kuiper belt objects \citep[e.g.][]{Ward(1976),Heppenheimer(1980),Ward(1981),NagasawaIda(2000),Nagasawa+(2000),Nagasawa+(2001),Nagasawa+(2002),Hahn(2003),Zheng+(2017)}.\footnote{Of course, direct gravitational scatterings (close encounters) can also play a role;
see, e.g.,~\cite{Nesvorny(2018)} and \cite{Broz+(2021)}.}
Extrasolar versions of this scenario have also been invoked to explain warm Jupiter eccentricities 
\citep[e.g.][]{Nagasawa+(2003),Petrovich+(2019),TeyssandierLai(2019)} and spin-orbit misalignments of hot Jupiters \citep[e.g.][]{LubowMartin(2016),Martin+(2016),SpaldingBatygin(2017),Vick+(2023)} and sub-Neptunes \citep{Petrovich+(2020), Epstein-Martin+(2022)}. \cite{Petrovich+(2020)} showed how a secular resonance between a gas giant and an interior sub-Neptune, both embedded in a decaying disc, could lift the sub-Neptune onto a polar orbit ($i_{\rm mut} \approx 90^\circ$).

We extend the works of \cite{Petrovich+(2020)} and others by studying how mutual inclinations of giant planets embedded in decaying transitional discs can be amplified by secular resonance passage. Our motivation includes misaligned giant planet systems like Kepler-448 and Kepler-693 \citep{Masuda(2017)}, and the proposed class of mutually inclined, apsidally orthogonal warm Jupiters (\citealt{DawsonChiang(2014)}). We also investigate how an inner disc can be tilted out of the plane of an outer disc by the action of an intermediary, gap-opening planet. Here we build upon the disc-tilting calculations of \citet{OwenLai(2017)} to study the parameter space occupied by transition discs shadowed by low-mass inner discs \citep{FrancisvanderMarel(2020),vanderMarel+(2021)}. Section~\ref{sec:InnerPlanetDisk} lays out the basic theory for how a secular resonance crossing can excite the inclination and eccentricity of an inner warm Jupiter (section \ref{sec:InnerPlanet}), and the inclination of an inner disc (section \ref{sec:InnerDisk}). Section~\ref{sec:Ext} revisits the \citet{DawsonChiang(2014)} hypothesis and updates the observed distribution of apsidal angles to see what might be inferred about mutual inclinations (section \ref{sec:Extra_OrthOrb}), and then examines how introducing the stellar spin and mass quadrupole affects our general results (section \ref{sec:Extra_SOMis}). Section~\ref{sec:SummaryDiscussion} summarizes and connects further with observations.

\section{Inclination and Eccentricity Excitation from Outer Disc Mass-Loss}
\label{sec:InnerPlanetDisk}

Consider a planet inside the cavity of a disc. The planet's longitude of
pericenter $\pom$ and longitude of ascending node $\Om$ precess at
rates proportional to the disc's mass.  As the disc's mass decreases,
the magnitudes of the precession frequencies $|\dot \pom|$ and
$|\dot \Om|$ decrease as well. If multiple bodies reside inside the cavity --- a pair
of planets, or a planet and an inner disc --- their
precession frequencies can be tuned down such that their respective apsidal or
nodal longitudes align, or anti-align, for extended periods of
time. As the bodies pass through such symmetric orbital configurations, angular momentum is transferred efficiently between 
them, with potentially large changes in eccentricity and inclination.

\subsection{Two Planets in a Transition Disc}
\label{sec:InnerPlanet}

\begin{figure*}
    \includegraphics[width=0.9\linewidth]{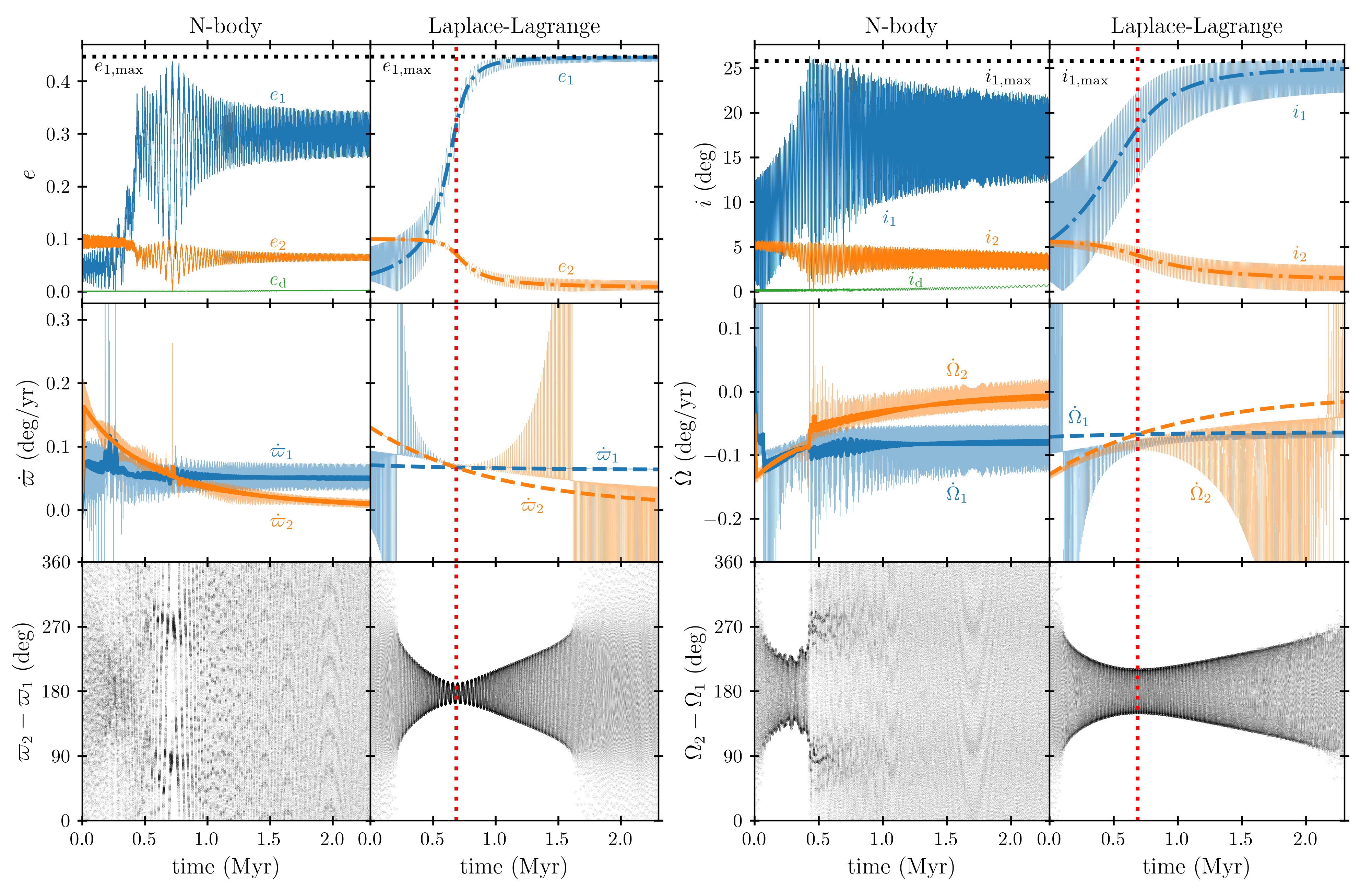}
    \caption{
    Secular resonance crossing of two planets inside a decaying transition disc, computed with the N-body code
   \texttt{REBOUND} \citep{ReinLiu(2012)}, and separately with Laplace-Lagrange theory (eqs.~\ref{eq:de1dt_LL}-\ref{eq:dOm2dt_LL}). 
   The inner planet's parameters are $\{m_1, a_1, e_{10}, i_{10}\} = \{1 \, \Mjup, 0.8 \, \au, 0.02, 0.02\}$ (with subscript 0 denoting initial conditions, 
    and the reference plane equal to the initial plane of the transition disc), and the outer planet's parameters are $\{m_2, a_2, e_{20}, i_{20}\} = \{10 \, \Mjup, 3.2 \, \au, 0.1, 0.1\}$. Modeled as a point mass in the N-body simulation, the disc has $\{m_{\rm d0}, a_{\rm d0}\} = \{0.22 \, \Msun, 14.3 \, \au\}$, while in our Laplace-Lagrange treatment, the disc has $\{m_{\rm d0}, \rin, \rout\} = (0.1 \, \Msun, 4.2 \, \au, 150 \, \au)$; these parameters yield similar times of resonance crossing, when apsides and nodes are anti-aligned ($\Dg \pom \approx \Dg \Om \approx 180^\circ$). As computed using Laplace-Lagrange theory, the maximum eccentricity and inclination  (eqs.~\ref{eq:e1_max_LL} and \ref{eq:i1_max_LL}) of the inner planet are shown as black horizontal dotted lines,    while dot-dashed blue and orange lines are solutions from an analytic model for secular resonance passage assuming anti-aligned apses and nodes  (the ``$-$'' branches of eqs.~\ref{eq:e1e2_lock} and \ref{eq:s1s2_lock}).    
    For the N-body data on $\dot \varpi$ and $\dot \Omega$, dark blue and orange 
    curves are time-averaged over a moving window of duration 18 kyr, while for the corresponding Laplace-Lagrange data, dashed curves are non-oscillatory contributions to precession frequencies (eqs.~\ref{eq:dpom1dt_sec}--\ref{eq:dOm2dt_sec}). Vertical dotted red lines mark when these non-oscillatory frequencies match and the secular resonance is crossed.
   }
    \label{fig:Nbody_LL_comp}
\end{figure*}

We consider two planets surrounded by an outer disc. The disc's mass
is prescribed to decrease with time. We compute the dynamical
evolution in two ways, first using the Laplace-Lagrange secular
equations (section \ref{subsec:LL}), and then with an $N$-body simulation (sections \ref{subsec:N}--\ref{subsec:NN}).

\subsubsection{Laplace-Lagrange theory}\label{subsec:LL}

Planet eccentricities $e$, pericenter longitudes
$\pom$, inclinations $i$, and nodal longitudes $\Om$ evolve according
to 
\begin{align}
\frac{\der e_1}{\der t} &= e_2 g_{12} \sin \Dg \pom
\label{eq:de1dt_LL} \\
\frac{\der e_2}{\der t} &= -e_1 g_{21} \sin\Dg \pom
\label{eq:de2dt_LL} \\
\frac{\der \pom_1}{\der t} &= (f_{12} + f_{1 \der}) - g_{12} \frac{e_2}{e_1} \cos \Dg \pom
\label{eq:dpom1dt_LL} \\
\frac{\der \pom_2}{\der t} &= (f_{21} + f_{2 \der}) - g_{21} \frac{e_1}{e_2} \cos \Dg \pom
\label{eq:dpom2dt_LL} \\
\frac{\der s_1}{\der t} &= -f_{12} s_2 \sin \Dg \Om
\label{eq:ds1dt_LL} \\
\frac{\der s_2}{\der t} &= f_{21} s_1 \sin \Dg \Om
\label{eq:ds2dt_LL} \\
\frac{\der \Om_1}{\der t} &= -(f_{12} + f_{1 \der}) + f_{12} \frac{s_2}{s_1} \cos \Dg \Om
\label{eq:dOm1dt_LL} \\
\frac{\der \Om_2}{\der t} &= -(f_{21} + f_{2 \der}) + f_{21}
                            \frac{s_1}{s_2} \cos \Dg \Om
\label{eq:dOm2dt_LL}
\end{align}
where subscripts 1 and 2 denote the inner and outer planet,
$\Delta \pom = \pom_2 - \pom_1$, $\Delta \Om = \Om_2 - \Om_1$, 
and $s = 2 \sin \frac{1}{2} i$. The planet-induced precession frequencies are
\begin{align}
f_{12} &= \frac{G m_1 m_2 a_1}{a_2^2 L_1} b_{3/2}^{(1)} \left(
           \frac{a_1}{a_2} \right) 
\label{eq:om12} \\
g_{12} &= \frac{G m_1 m_2 a_1}{a_2^2 L_1} b_{3/2}^{(2)} \left(
           \frac{a_1}{a_2} \right)
\label{eq:nu12} \\
f_{21} &= \frac{L_1}{L_2} f_{12}, \hspace{5mm} g_{21} = \frac{L_1}{L_2} g_{12}, \label{eq:back-reac}
\end{align}
for gravitational constant $G$, planet mass $m$, semi-major axis $a$
(conserved in this secular theory),
leading-order angular momentum 
$L_k = m_k \sqrt{G M_\star a}$, stellar mass $M_\star$, and Laplace
coefficient $b$ (\citealt{MurrayDermott(2000)}; \citealt{PuLai(2018)}).
The disc-induced precession frequency of planet $k$ is
\begin{equation}
f_{k\der} = \frac{G m_k a_k}{L_k} \int_{\rin}^{\rout} \frac{2\pi
  \Sg(r)}{r} b_{3/2}^{(1)} \left( \frac{a_k}{r} \right) \der r 
\label{eq:omkd}
\end{equation}
where we assume the disc's surface density profile follows 
\begin{equation}
\Sg(t,r) = \frac{\md(t)}{\pi \rout r} 
\end{equation}
with a disc mass that decays exponentially with time
\begin{equation}
m_{\rm d}(t) = m_{\rm d0} \e^{-t/t_{\rm d}} \,.
\end{equation}
Note that for now we do not include the back-reaction of the
planets onto the disc; this restriction is relaxed in our $N$-body
calculation in section \ref{subsec:N}. The equations are solved using the \texttt{scipy.integrate.odeint} integrator in \texttt{python}.

Figure~\ref{fig:Nbody_LL_comp} displays an example evolution
for two giant planets with $m_1 = 1 \ {\rm m}_{\rm J}$ and $m_2 = 10\  {\rm m}_{\rm J}$, located at $a_1 = 0.8$ au and $a_2 = 3.2$ au from a
star of mass $M_\star = {\rm M}_\odot$. Initial eccentricities and
inclinations are $e_{10}, i_{10} = 0.02$ and $e_{20}, i_{20} =
0.1$, and initial longitudes are $\pom_{10}, \pom_{20}, \Om_{10}, \Om_{20} = 0$. For the disc, $m_{\rm d0} = 0.1 \ 
  {\rm M}_\odot$, $r_{\rm in} = 4.2$ au, 
$r_{\rm out} = 150$ au, 
and $t_{\rm d} =
1$ Myr.  Such a $m_{\rm d0}$ value is comparable to disc mass estimates for Class 0/I sources \citep[e.g.][]{Jorgensen+(2009), Tobin+(2015), Segura-Cox+(2018), Andersen+(2019)}, and necessary for a secular resonance crossing for our giant planet parameters.
Over the course of the integration, as disc mass $m_{\rm d}$ decreases, we see $e_1$ and $i_1$  amplify at the
expense of $e_2$ and $i_2$. The changes are fastest when $\Dg \pom
\approx \pi$ and $\Dg \Om \approx \pi$. The apsidal and nodal
anti-alignments are 
prolonged by the matching of frequencies $\dot \pom_1$ with
$\dot \pom_2$, and $\dot \Om_1$ with $\dot \Om_2$.

Looking more closely at these precession frequencies, we see from
equations~\eqref{eq:dpom1dt_LL}-\eqref{eq:dpom2dt_LL}
and~\eqref{eq:dOm1dt_LL}-\eqref{eq:dOm2dt_LL} that they are each
composed of two terms.  One contribution to the frequency depends on the degree of
misalignment, either $\Dg \pom$ or $\Dg \Om$; this planet-planet
interaction frequency oscillates rapidly
and attains large values, positive or negative, whenever 
eccentricity or inclination become small. By contrast, the other non-oscillatory frequencies
\begin{align} 
\left. \frac{\der \pom_1}{\der t} \right|_{\rm non-osc} &= f_{12} +
                                                          f_{1
                                                          \der} 
\label{eq:dpom1dt_sec}  
\\ 
\left. \frac{\der \pom_2}{\der t} \right|_{\rm non-osc} &= f_{21} + f_{2 \der}
\label{eq:dpom2dt_sec} 
\\
\left. \frac{\der \Om_1}{\der t} \right|_{\rm non-osc} &= -(f_{12} + f_{1 \der})
\label{eq:dOm1dt_sec} 
\\
\left. \frac{\der \Om_2}{\der t} \right|_{\rm non-osc} &= -(f_{21} + f_{2 \der})
\label{eq:dOm2dt_sec} 
\end{align}
change much more gradually with time as the outer disc loses mass. These slowly
varying frequencies, plotted as dashed curves in Fig.~\ref{fig:Nbody_LL_comp}, cross
at a time marked with a vertical red dotted line. At this moment of
``secular resonance crossing'', mean eccentricities and inclinations
change fastest.

We can place bounds on how much eccentricities and inclinations grow
by examining the constants of motion admitted by eqs.~\eqref{eq:de1dt_LL}-\eqref{eq:de2dt_LL} and~\eqref{eq:ds1dt_LL}-\eqref{eq:ds2dt_LL}:
\begin{align}
\frac{1}{2} L_1 e_1^2 + \frac{1}{2} L_2 e_2^2 &= \text{constant} \label{eq:e_const} \\
\frac{1}{2} L_1 s_1^2 + \frac{1}{2} L_2 s_2^2 &=
                                                \text{constant}. \label{eq:s_const} 
\end{align}
When the outer planet has much more angular momentum than the inner ($L_2 \gg L_1$,
as is the case in Fig.~\ref{fig:Nbody_LL_comp}), the maximum
eccentricity and inclination that can be attained by the inner planet depend on the outer planet's initial conditions:
\begin{align}
e_{\rm 1,max} &\simeq \sqrt{ \frac{L_2}{L_1} } e_{20} \label{eq:e1_max_LL} \\
i_{\rm 1,max} &\simeq 2 \sin^{-1} \left( \sqrt{ \frac{L_2}{L_1} } \sin
                \frac{1}{2} i_{20} \right) \,.  \label{eq:i1_max_LL}
\end{align}
Fig.~\ref{fig:Nbody_LL_comp} demonstrates that $e_1$ and $i_1$
come close to their respective maxima after the secular resonance crossing. Appendix~\ref{app:SweepSecRes} and section \ref{subsec:NN} explore in more detail why $L_2 > L_1$ and passage through anti-aligned states are preferred for inclination and eccentricity excitation.

\subsubsection{N-body}\label{subsec:N}
We solve the same problem as above (two planets inside a
decaying outer disc) now using the \texttt{REBOUND} $N$-body code outfitted with the \texttt{IAS15} integrator
\citep{ReinSpiegel(2015)}. The outer disc is modeled as a point particle whose
mass $m_{\rm d}$ decays exponentially with time constant $t_{\rm d} =
1$ Myr; to effect this, we use the \texttt{REBOUNDx} routine
\texttt{modify\_mass} (\citealt{Kostov+(2016),Tamayo+(2020)}) setting $\texttt{mass\_loss} = -t_{\rm d}$. The initial disc mass of $0.22 \ {\rm M}_\odot$ and initial semi-major axis of $a_{\rm d} = 14.3$ au are
chosen to yield a secular resonance crossing time comparable to that
in our Laplace-Lagrange solution above ($\sim$0.7 Myr).
Modeled as a point particle just like the planets, the disc has an eccentricity
$e_{\rm d}$ and inclination $i_{\rm d}$ that are free to vary; in
practice they do not deviate much from their initial values of $0$, as
the disc contains the lion's share of the system's angular momentum (this assumption eventually breaks down as $m_\der$ decreases to zero, resulting in a slight increase in $e_\der$ and $i_\der$ at late times $t\gtrsim 5 \, \Myr$, not plotted).

Fig.~\ref{fig:Nbody_LL_comp} shows that the $N$-body solution
broadly matches the Laplace-Lagrange solution --- $e_1$ and $i_1$ increase
while $e_2$ and $i_2$ decrease as a result of a secular resonance crossing. In the $N$-body evolution, the
magnitude of the changes in eccentricity and inclinaton are more
muted, apsidal and nodal anti-alignments
between planets 1 and 2 are more short-lived, and precession frequencies match only
momentarily and not at all after the secular resonance is crossed.

By modelling the disc as a particle in our $N$-body simulation, we neglect how planets excite and interact with waves in the disc \citep[e.g.][]{GoldreichTremaine(1980), TanakaWard(2004)}. Most relevant to our study are long-wavelength apsidal and nodal waves which can exchange angular momentum with a planet \citep[][]{GoldreichSari(2003)}. Our $N$-body particle treatment does allow for angular momentum exchange between the disc and planets, but neglects how that exchange depends on the wave nature of apsidal and nodal disturbances in the disc.

\subsubsection{Parameter space exploration ($N$-body)}\label{subsec:NN}

\begin{figure}
   \includegraphics[width=\columnwidth]{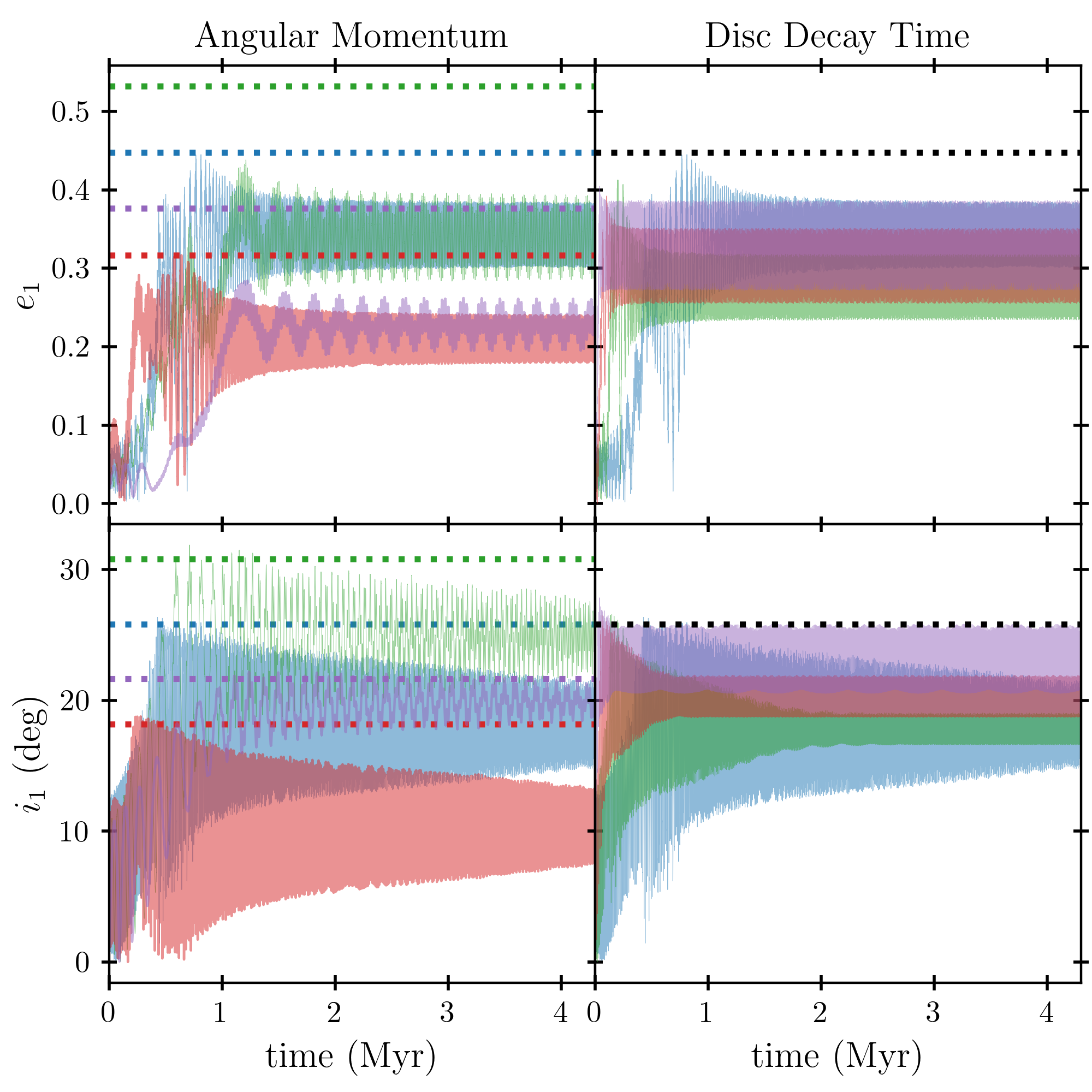}
    \caption{Parameter space exploration for simulations with two planets + a decaying outer disc. All runs shown here use \texttt{REBOUND}, with $\{m_1, a_1\} = \{1 \, \Mjup, 0.8 \, \au\}$, $e_{10} = i_{10} = 0.02$, and $e_{20} = i_{20} = 0.1$, the same parameters as in Fig.~\ref{fig:Nbody_LL_comp}. \textit{Left panels}: Effects of varying the angular momentum of the outer planet (and concomitantly the outer disc). In order of increasing angular momentum: \{$m_2$, $a_2$, $\mdo$, $\ado$\} = \{$5 \, \Mjup$, $3.2 \, \au$, $0.20 \, \Msun$, $17.2\, \au$\} (red), \{$5 \, \Mjup$, $6.4 \, \au$, $0.12 \, \Msun$, $42.8 \, \au$\} (purple), \{$10 \, \Mjup$, $3.2 \, \au$, $0.22 \, \Msun$, $14.3 \, \au$\} (blue), and \{$10 \, \Mjup$, $6.4 \, \au$, $0.14 \, \Msun$, $35.3 \, \au$\} (green), with $t_{\rm d} = 1 \, \Myr$. Dotted lines denote $e_{\rm 1,max}$ and $i_{\rm 1,max}$ as computed using eqs.~\eqref{eq:e1_max_LL} and \eqref{eq:i1_max_LL}. \textit{Right panels}: Effects of varying the disc decay time. In order of decreasing decay time:  $t_{\rm d} = 1 \, \Myr$ (blue), $0.3 \, \Myr$ (green), $0.1 \, \Myr$ (red), and $0.03 \, \Myr$ (purple), with $\{m_2, a_2, \mdo, \ado\} = \{10 \, \Mjup, 3.2 \, \au, 0.22 \, \Msun, 14.3 \, \au\}$.}
 \label{fig:Res_pars}
\end{figure}

In Figure~\ref{fig:Res_pars}, we explore how changing the angular
momenta of the outer planet + disc (while keeping $L_2 > L_1$), and separately the disc depletion
time, affect the eccentricity and inclination evolution of the inner planet in our $N$-body runs. As the outer bodies' angular momenta are increased, the final $i_1$ increases as well, respecting $i_{\rm 1,max}$. A similar trend plays out for $e_1$. We also see that the excitation of $e_1$ and $i_1$ do not much depend on $t_{\rm d}$, presumably as long as the decay time exceeds the secular oscillation periods (e.g. $f_{12}^{-1}, f_{2d}^{-1}$). Note, however, how the shortest decay times sampled in Fig.~\ref{fig:Res_pars} yield the largest $i_1$.

\begin{figure*}
    \centering  \includegraphics[width=0.81\linewidth]{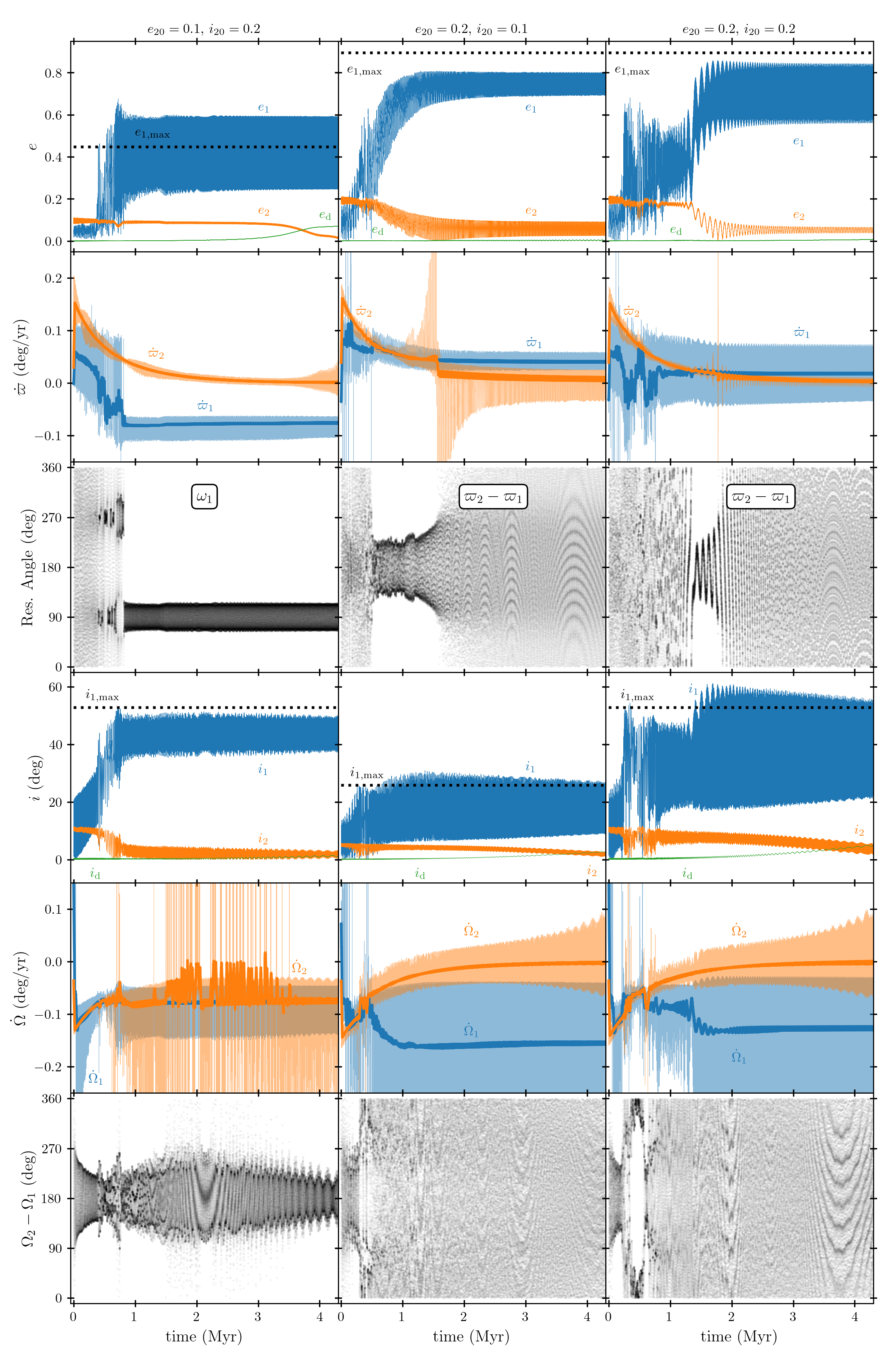}
   \caption{Similar to Fig.~\ref{fig:Nbody_LL_comp}, showing N-body simulations of two planets + a decaying outer disc, and varying the initial eccentricity $e_{20}$ and initial inclination $i_{20}$ of the outer planet as indicated above each column. Remaining planet and disc parameters are $\{m_1, a_1, m_2, a_2, \mdo, \ado\} = \{1 \, \Mjup, 0.8 \, \au, 10 \, \Mjup, 3.2 \, \au, 0.22 \, \Msun, 14.3 \, \au\}$ and $i_{10} = e_{10} = 0.02$. In the left column the system captures into the Lidov-Kozai resonance where the argument of pericentre $\omega_1 = \varpi_1 - \Omega_1$ librates about 90$^\circ$, and $e_1$ does not respect $e_{\rm 1,max}$ as given by \eqref{eq:e1_max_LL} from Laplace-Lagrange theory.}
\label{fig:Nbody_pars}
\end{figure*}

Figure~\ref{fig:Nbody_pars} explores the evolution when the outer
planet's initial eccentricity $e_{20}$ and inclination $i_{20}$ are set to higher 
values. Outcomes are largely the same as before, except for the run
where $e_{20} < i_{20}$ (first column), where an eccentricity secular resonance is
not encountered, and the final $e_1$ does not respect 
$e_{\rm 1,max}$ as derived from Laplace-Lagrange. What happens instead
is that $i_1$ grows to values $\gtrsim 40^\circ$, large enough for
$e_1$ to trade off with $i_1$ in a Lidov-Kozai resonance
(not captured by Laplace-Lagrange) --- see how the periastron argument
$\omega_1$ starts to librate near the Kozai fixed points of $\pm$90$^\circ$ and eventually locks onto 90$^\circ$.

\begin{figure*}
    \centering \includegraphics[width=0.9\linewidth]{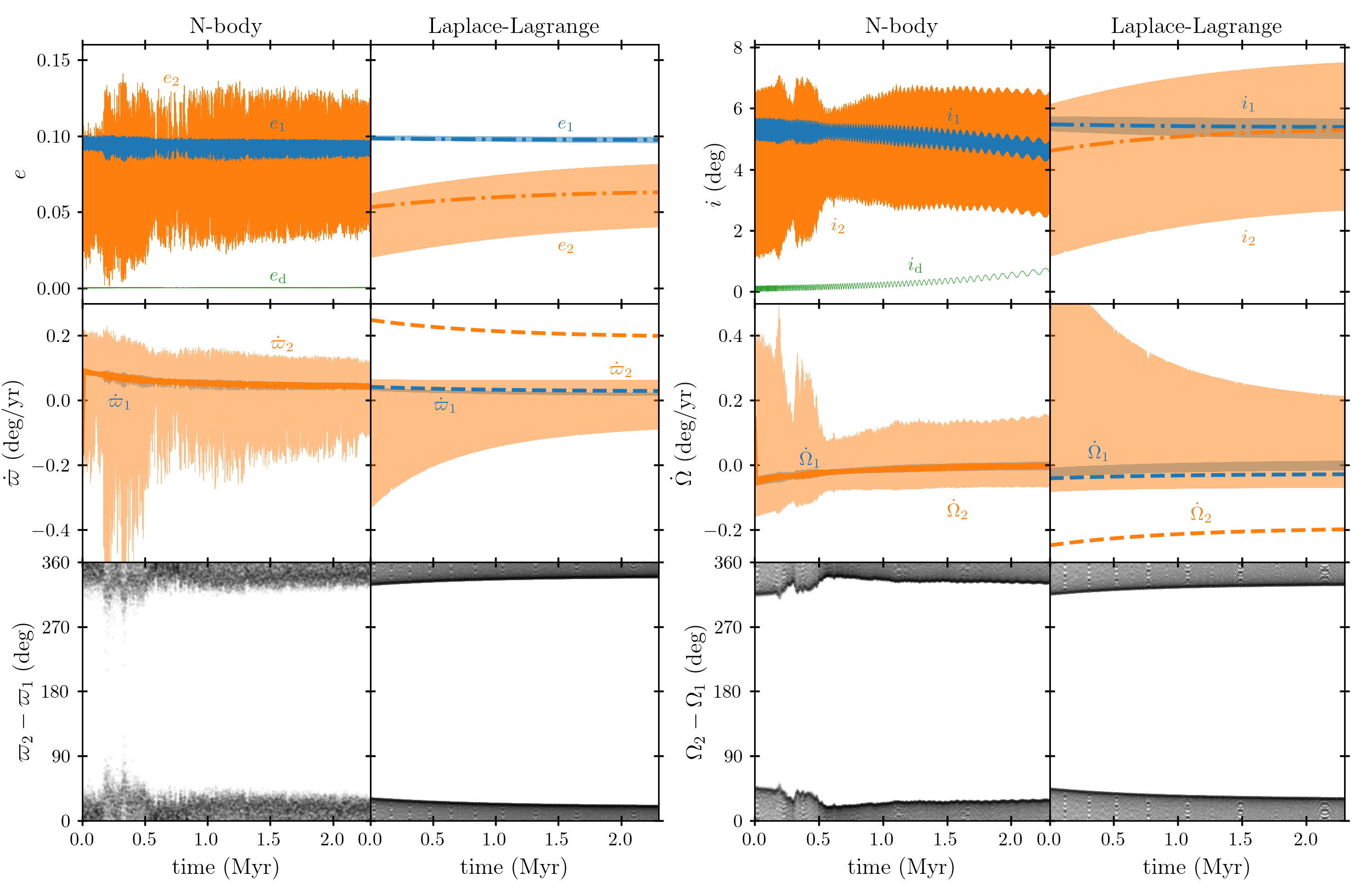}
   \caption{
   Similar to Fig.~\ref{fig:Nbody_LL_comp}, showing simulations of two planets + a decaying outer disc, but now with the outer planet having less angular momentum than the inner: \{$m_1$, $a_1$, $m_2$, $a_2$\} = \{$10 \, \Mjup$, $1.6 \, \au$, $1 \, \Mjup$, $3.2 \, \au$\}. Initial eccentricities and inclinations are $e_{10} = i_{10} = 0.1$ and $e_{20} = i_{20} = 0.02$, and disc parameters are identical to those in Fig.~\ref{fig:Nbody_LL_comp}.    In the Laplace-Lagrange solution, the non-oscillatory precession frequencies of the two planets never match (dashed curves; eqs.~\ref{eq:dpom1dt_sec}-\ref{eq:dOm2dt_sec}), no secular resonance is crossed, and eccentricities and inclinations do not undergo large changes aside from the outer planet's secular oscillations. A qualitatively similar evolution plays out in the $N$-body calculation. Note how the planets are apsidally and nodally aligned from start to finish but do not much change their eccentricities or inclinations; see Appendix \ref{app:SweepSecRes}.
   }
\label{fig:Nbody_lowL}
\end{figure*}

We have focused so far on the case where the outer planet has more angular momentum than the inner. We find that when the angular momentum ratio is flipped, the planets do not cross a secular resonance. When $L_1 > L_2$, the planet-planet precession frequencies $f_{12} < f_{21}$ (eq.~\ref{eq:back-reac}). Meanwhile the planet-disc frequencies satisfy $f_{1\der} < f_{2\der}$ since the exterior planet lies closer to the cavity edge (eq.~\ref{eq:omkd}).  Hence the magnitudes of the non-oscillatory nodal and apsidal precession frequencies of the outer planet, $f_{21}+f_{2\der}$, always stay higher than those of the inner planet, $f_{12}+f_{1\der}$, and there is never a crossing (eqs.~\ref{eq:dpom1dt_sec}-\ref{eq:dOm2dt_sec}).  Figure~\ref{fig:Nbody_lowL} plots a sample $L_2 < L_1$ integration where $\{m_1, a_1\} = \{10 \, \Mjup, 1.6 \, \au\}$ and $\{m_2, a_2\} = \{1 \, \Mjup, 3.2 \ \au\}$.  The system does not pass into or out of an aligned or anti-aligned state; in this particular example, it starts and stays within an aligned state, and the planet eccentricities and inclinations do not change much. Appendix~\ref{app:SweepSecRes} explains this result in greater detail.

\subsection{Inner Disc and Planet in a Transition Disc Cavity}
\label{sec:InnerDisk}

\begin{figure*}
    \centering
    \includegraphics[width=0.9\linewidth]{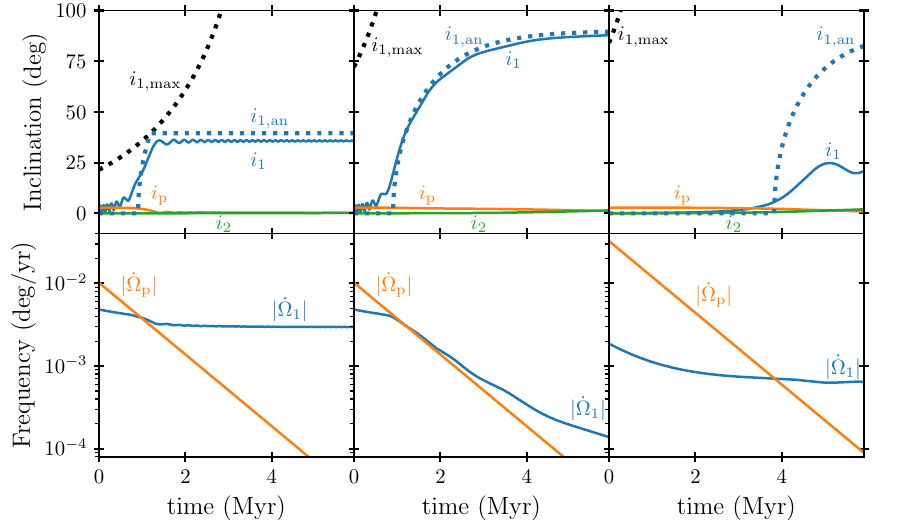}
    \caption{ 
    Secular evolution of a planet (subscript p) sandwiched between a decaying inner disc (subscript 1) and decaying exterior disc (subscript 2), calculated using the vector equations of section \ref{sec:InnerDisk}. 
    Parameters are $\{t_{\rm d}, m_{20}, m_{\rm p}, r_{\rm 1,in}, r_{\rm 1,out}, r_{\rm 2,in}, r_{\rm 2,out}\} = \{1 \, {\rm Myr}, 0.2 \, {\rm M}_\odot, 5 \, \Mjup, 0.03 \, \au, 5\, \au, 30 \ \au, 200 \, \au\}$, with each column corresponding to $\{a_{\rm p}, \delta\} = \{12 \, \au, 2 \times 10^{-2}\}$ (left), $\{12 \, \au, 2 \times 10^{-3}\}$ (middle), and $\{20 \, \au, 2 \times 10^{-3}\}$ (right), respectively, where $\delta \equiv (m_1/m_2) (r_{\rm 2,out}/r_{\rm 1,out})$ (equation \ref{eq:M_disks}; see also fig.~9 of \citealt{FrancisvanderMarel(2020)} for observationally inferred values of $\delta$). 
    The analytic prediction for the inner disc inclination $i_{\rm 1,an}$ (eq.~\ref{eq:i1_an} and text below eq.~\ref{eq:i1_max_d}) agrees well with the numerically computed $i_1$, except for the simulation in the right column where secular resonance ($\dot \Om_1 \approx \dot \Om_{\rm p}$) cannot be maintained because the planet's angular momentum exceeds that of the outer disc.   
    }
    \label{fig:InnerDiskTilt}
\end{figure*}

Here we replace the inner planet of section \ref{sec:InnerPlanet} with
a disc. We now have two discs, one interior to a planet, and another
exterior. Rigid precession of either disc is assumed to be enforced by 
some collective effect, e.g.~bending waves \citep[e.g.][]{LubowOgilvie(2000),ZanazziLai(2018a),Nealon+(2018),Zhu(2019)} or disc self-gravity \citep[e.g.][]{ZanazziLai(2017),Batygin(2018)}. As with the outer
disc, the inner disc loses mass; the inner disc's angular momentum can
become small relative to the planet's, enabling the excitation of
large mutual inclination (equation \ref{eq:i1_max_LL}). To more accurately model large inclinations,
we replace the Laplace-Lagrange equations of section \ref{subsec:LL}
with the vector formalism of \cite{LaiPu(2017)}, which accounts for
how nodal precession rates depend on inclination (see also the
appendix of \citealt{ZanazziLai(2017)}). These vector equations are designed to model well-separated masses, and are thus
appropriately applied to large transition
disc cavities far removed from their inner discs \citep[e.g.][]{FrancisvanderMarel(2020),Bohn+(2022)}. For simplicity we fix the
eccentricities of all masses to be zero, but account for back-reaction
by allowing the outer disc inclination to freely evolve.

The orbit normals of the inner disc ($\hl_1$), planet ($\hlpl$), and outer disc
($\hl_2$) obey:
\begin{align}
\frac{\der \hl_1}{\der t} &= f_{\rm 1p} (\hl_1 \bcdot \hlpl)(\hl_1
                            \btimes \hlpl) + f_{\rm 12}(\hl_1 \bcdot
                            \hl_2)(\hl_1 \btimes \hl_2) \label{eq:dhl1dt} \\
\frac{\der \hlpl}{\der t} &= f_{\rm p1} (\hlpl \bcdot \hl_1)(\hlpl \btimes \hl_1) + f_{\rm p2} (\hlpl \bcdot \hl_2)(\hlpl \btimes \hl_2) \label{eq:dhlpldt} \\
\frac{\der \hl_2}{\der t} &= f_{21} (\hl_2 \bcdot \hl_1)(\hl_2 \btimes
                            \hl_1) + f_{\rm 2p} (\hl_2 \bcdot
                            \hlpl)(\hl_2 \btimes \hlpl) \,. \label{eq:dhl2dt}
\end{align}
The precession frequencies are given by:
\begin{align}
f_{\rm 1p} &= \frac{1}{L_1} \int_{\roin}^{\roout} \frac{2\pi G \Sigma_1 m_{\rm p} r_1^2}{a_{\rm p}^2} b_{3/2}^{(1)} \left( \frac{r_1}{a_{\rm p}} \right) \der r_1
\label{eq:f1p} \\
f_{\rm p2} &= \frac{1}{L_{\rm p}} \int_{\rtin}^{\rtout} \frac{2\pi G \Sigma_2 m_{\rm p} a_{\rm p}}{r_2} b_{3/2}^{(1)} \left( \frac{\apl}{r_2} \right) \der r_2
\label{eq:fp2} \\
f_{12} &= \frac{1}{L_1} \int_{\rtin}^{\rtout} \int_{\roin}^{\roout} \frac{4\pi^2 G \Sigma_1 \Sigma_2 r_1^2}{r_2} b_{3/2}^{(1)} \left( \frac{r_1}{r_2} \right) \der r_1 \der r_2
\label{eq:f12} \\
f_{\rm p1} &= \frac{L_1}{L_{\rm p}} f_{\rm 1p}, \hspace{3mm}
f_{\rm 2p} = \frac{L_{\rm p}}{L_2} f_{\rm p2}, \hspace{3mm}
f_{\rm 21} = \frac{L_1}{L_2} f_{12}
\end{align}
where the inner disc of radial coordinate $r_1$ 
extends from  $\roin$ 
to $\roout$, and similarly for the outer disc. The planet's angular
momentum (to leading order) is $L_{\rm p} = m_{\rm p} \sqrt{G M_\star
  a_{\rm p}}$ for planet mass $m_{\rm p}$ and semi-major axis $a_{\rm
  p}$, stellar mass $M_\star$, and gravitational constant $G$. The
inner and outer disc surface density profiles are assumed to follow:
\begin{equation}
    \Sigma_1(t,r_1) = \frac{m_1(t)}{2\pi r_{\rm 1,out} r_1},
    \hspace{3mm}
    \Sigma_2(t,r_2) = \frac{m_2(t)}{2\pi r_{\rm 2,out} r_2}
    \label{eq:Sigma_disks}
\end{equation}
for disc masses
\begin{align}
    m_1(t) = \delta \frac{r_{\rm 1,out}}{r_{\rm 2,out}} 
             m_2(t), \hspace{2mm}
    m_2(t)  = m_{20} \e^{-t/t_{\rm d}} \label{eq:M_disks} 
\end{align}
where $\dg$ is a free parameter that measures how much lower the inner disc surface density is relative to the outer (where the latter is extrapolated to the same inner disc radius; see section 3.3 and fig.~5 of \citealt{FrancisvanderMarel(2020)}). For $r_{\rm 1,in} \ll r_{\rm 1,out}$ and $r_{\rm 2,in} \ll r_{\rm
  2,out}$, the (leading-order) disc angular momenta are:
\begin{align}
    L_1(t) &= \frac{2}{3} m_1(t) \sqrt{G \ms r_{\rm 1,out}} \\
    L_2(t) &= \frac{2}{3} m_2(t) \sqrt{G \ms r_{\rm 2,out}} \,.
\end{align} 

Figure~\ref{fig:InnerDiskTilt} plots three sample integrations of equations~\eqref{eq:dhl1dt}-\eqref{eq:dhl2dt} for different choices of $a_{\rm p} = \{12, 20\}$ au and $\delta = \{2 \times 10^{-3}, 2 \times 10^{-2}\}$, values motivated by observations of material inside the cavities of transitional discs \citep[e.g.][]{UbeiraGabellini+(2019),FrancisvanderMarel(2020),Portilla-Revelo+(2023)}. The annular extent of the inner disc is taken to be $\{r_{\rm 1,in}, r_{\rm 1,out}\} = \{0.03, 5\} \, \au$, and for the outer disc $\{r_{\rm 2,in}, r_{\rm 2,out}\} = \{30, 200\} \, \au$.
In all cases the inner and outer discs are assumed initially co-planar, and the planet has an initial seed inclination of $i_{\rm p0} = 3^\circ$. We see that $i_1$ amplifies when the nodal precession rate of the inner disc:
\begin{align}
  \dot \Om_{\rm 1} \simeq -f_{\rm 1p} \cos i_1 - f_{\rm 12} \cos i_1
  \label{eq:dOm1} 
  \end{align}
  matches that of the planet:
  \begin{align}
  \dot \Om_{\rm p} \simeq -f_{\rm p1} \cos i_1 - f_{\rm p2}
  \label{eq:dOmp}
\end{align}
where we have taken the reference plane to be perpendicular to the initial $\hl_2$, approximated $\hlpl$ and $\hl_2$ to be constant and nearly parallel (as Fig.~\ref{fig:InnerDiskTilt} confirms), and kept only the non-oscillatory contributions to the frequencies (dropping the dependence of $\dot \Om_1$ and $\dot \Om_{\rm p}$ on the relative nodes). The nodal frequencies $\dot \Om_1$ and $\dot \Om_{\rm p}$ can track one another for some time; see especially the middle column of Fig.~\ref{fig:InnerDiskTilt}. 
Setting $\dot \Om_1 = \dot \Om_{\rm p}$ gives an analytic estimate for $i_1$ during resonance lock:
\begin{equation}
    \cos i_{\rm 1,an} \approx \frac{f_{\rm 2p}}{f_{\rm 1p} - f_{\rm p1} + f_{12}}.
    \label{eq:i1_an}
\end{equation}
As $f_{\rm 2p} \to 0$, $i_{\rm 1,an} \to 90^\circ$ \citep{Petrovich+(2020)}. 

We can also derive a maximum $i_1$ as we did in section \ref{sec:InnerPlanet}. From equations \eqref{eq:dhl1dt} and \eqref{eq:dhlpldt} and $\der \hl_2/\der t = 0$ (back-reaction on the outer disc neglected), we have 
\begin{equation}
L_1 (1 - \cos i_1) + L_{\rm p} (1 - \cos i_{\rm p}) \simeq \text{constant}
\end{equation}
which yields 
\begin{equation}
i_{\rm 1,max} \simeq 2 \sin^{-1} \left( \sqrt{ \frac{L_{\rm p}}{L_1} } \sin \frac{1}{2} i_{\rm p0} \right)
\label{eq:i1_max_d}
\end{equation}
when $L_1 \ll L_{\rm p}$. This result is identical to equation \eqref{eq:i1_max_LL} as derived using Laplace-Lagrange.

Equation~\eqref{eq:i1_an} holds when $f_{\rm 2p} < f_{\rm 1p} - f_{\rm p1} + f_{12}$ and $i_{\rm 1,an} < i_{\rm 1, max}$. We can piece together a more complete analytic solution for $i_1$ by setting $i_{\rm 1,an} = 0$ when $f_{\rm 2p} \ge f_{\rm 1p} - f_{\rm p1} + f_{12}$, and further setting $i_{\rm 1,an}(t) = i_{\rm 1,max}(t_{\rm c})$ where $t_{\rm c}$ is the time when $i_{\rm 1,an}$ first crosses $i_{\rm 1,max}$. We see from Fig.~\ref{fig:InnerDiskTilt} that $i_{\rm 1,an}$ so constructed agrees well with the full solutions shown in the left and middle columns. If the planet's angular momentum exceeds that of the outer disc when secular resonance is first encountered (i.e.~if $L_{\rm p} \gtrsim L_2$ 
when $\dot \Om_1 \approx \dot \Om_p$), secular resonance locks cannot be sustained, and equation~\eqref{eq:i1_an} is a poor predictor of $i_1$ (right column).

\begin{figure}
    \centering
    \includegraphics[width=1.0\columnwidth]{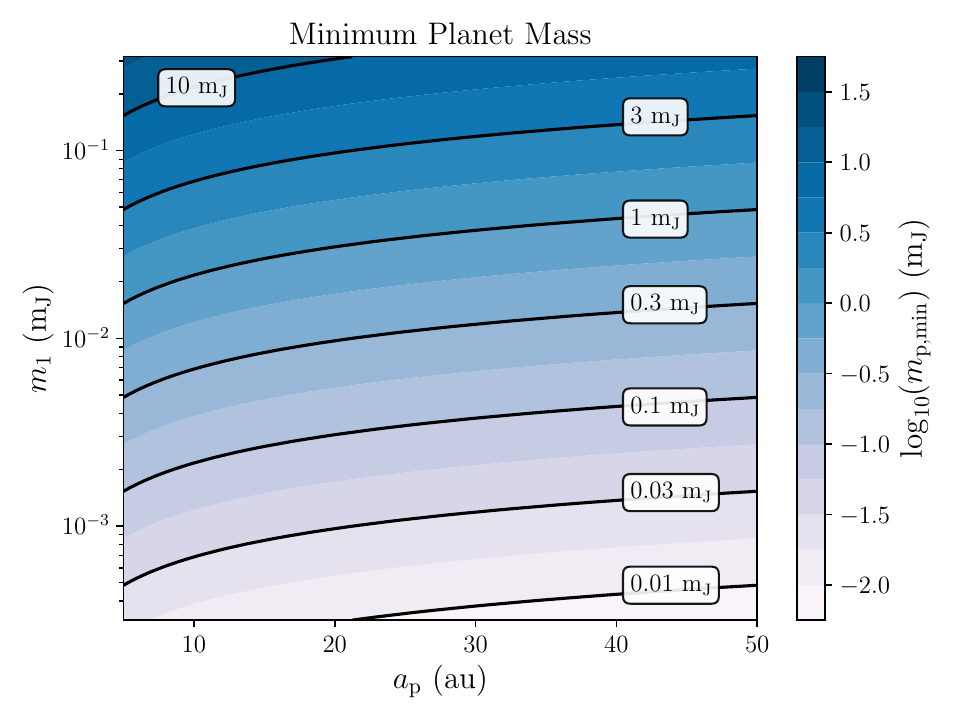}
    \caption{
    Minimum planet mass $m_{\rm p,min}$ required to tilt the inner disc by $i_{\rm 1,max} = 30^\circ$ (eq.~\ref{eq:mp_min}), as a function of the inner disc mass $m_1$ and the planet semimajor axis $a_{\rm p}$, 
    assuming $\{r_{\rm 1,out}, i_{\rm p0}\} = \{5 \, \au, 3^\circ\}$. See also Figure \ref{fig:CQTau} which shows that this minimum planet mass satisfies observational constraints in the transition disc hosted by CQ Tau.}
    \label{fig:Mp_min}
\end{figure}

One can rearrange equation~\eqref{eq:i1_max_d} to estimate the minimum planet mass required to tilt the inner disc:
\begin{equation}
m_{\rm p} > m_{\rm p,min} = \frac{2}{3} m_1 \left( \frac{\roout}{\apl} \right)^{1/2} \frac{\sin^2 \frac{1}{2} i_{\rm 1,max}}{\sin^2 \frac{1}{2} i_{\rm p0}}.
\label{eq:mp_min}
\end{equation}
Figure~\ref{fig:Mp_min} plots $m_{\rm p,min}$ as a function of $\apl$ and $m_1$.  Take for example an inner disc having mass $m_1 \gtrsim 0.01 \ {\rm m}_{\rm J}$ on scales of $\roout = 5$ au. A planet at $\apl = 10$-50 au with mass $m_{\rm p} \gtrsim 0.3 \ {\rm m}_{\rm J}$ can tilt such an inner disc from $3^\circ$ to $30^\circ$, provided they are surrounded by an outer disc dominating the angular momentum budget.

\section{Extensions}
\label{sec:Ext}

\subsection{Forming Inclined, Apsidally-Orthogonal Planetary Systems}
\label{sec:Extra_OrthOrb}

\begin{figure*}
    \centering
    \includegraphics[width=0.95\textwidth]{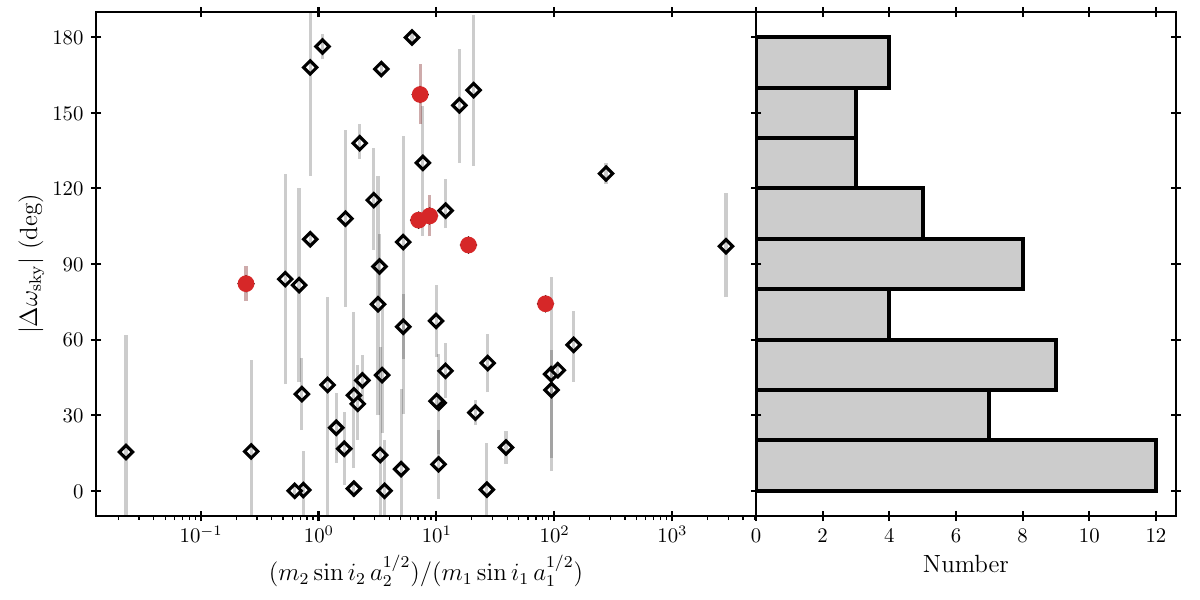}
    \caption{
    The degree of apsidal alignment in two-planet systems with radial velocity or transit timing data, as measured by the difference in sky-projected arguments of pericentre, $|\Delta \om_{\rm sky}| = |\om_{\rm sky,1} - \om_{\rm sky,2}|$. Data are taken from the NASA Exoplanet Archive, Exoplanet Encyclopedia, and Exoplanet Orbit Database, selecting only systems where 
    eccentricities are larger than zero with greater than two-sigma confidence, and one-sided one-sigma errors on $\om_{{\rm sky},}$ are less than $40^\circ$ (similar cuts were made by \protect\citealt{DawsonChiang(2014)}).     No cut on planet mass or semi-major axis is made for this figure, but we have verified that our conclusions are unchanged if we consider only giant planets, or if we include only planets with $a > 0.1 \ \au$ to exclude hot Jupiters (data not shown). The six red points are the same six warm Jupiter systems proposed by   \protect\cite{DawsonChiang(2014)} to constitute a distinct class of apsidally misaligned ($|\Delta \om_{\rm sky}| \approx 90^\circ$), highly inclined pairs. The data as a whole no longer support such a claim, as can be seen in the histogram of $|\Dg \om_{\rm sky}|$ on the right, which shows no obvious clustering around $|\Dg \om_{\rm sky}| \approx 90^\circ$, and instead appears more consistent with most if not all pairs having small mutual inclinations --- compare with Figure \ref{fig:Hist_PlanetPeri}.
    }
    \label{fig:Peri_Data}
\end{figure*}

\begin{figure}
    \centering  
    \includegraphics[width=1.0\linewidth]{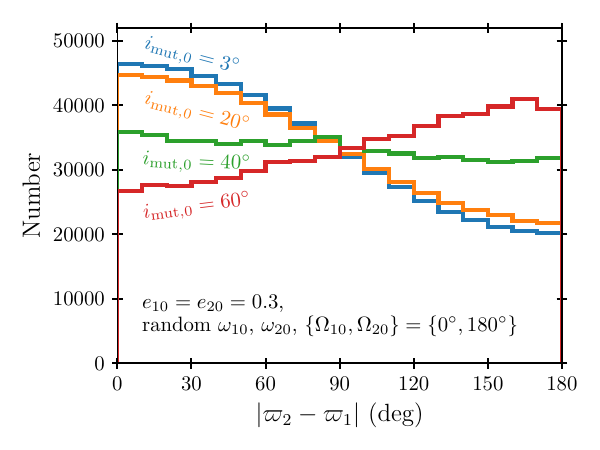}
   \caption{
   The distribution of $|\Dg \pom| = |\pom_2 - \pom_1|$  sampled from 200 two-planet systems, integrated using eqs.~(17)-(20) of \protect\cite{Liu+(2015)} over 1 Myr.  The inner planet has \{$m_1$, $a_1$\} = \{$1 \, \Mjup$, $0.2 \, \au$\} and the outer planet has \{$m_2$, $a_2$\} = \{$3 \, \Mjup$, $1.0 \, \au$\}. Initial mutual inclinations are as labeled, with initial nodes $\Omega_{10}$ and $\Omega_{20}$ anti-aligned, initial arguments of pericenter drawn randomly over the interval $[0,2\pi]$, and initial eccentricities set to 0.3. For $i_{\rm mut,0} \lesssim 20^\circ$, the $\Dg \pom$ distribution decreases monotonically from alignment to anti-alignment in a way that resembles the observed distribution of $|\Delta \omega_{\rm sky}|$ shown in Figure \ref{fig:Peri_Data}. 
   }
\label{fig:Hist_PlanetPeri}
\end{figure}

\begin{figure}
    \centering  
    \includegraphics[width=0.9\linewidth]{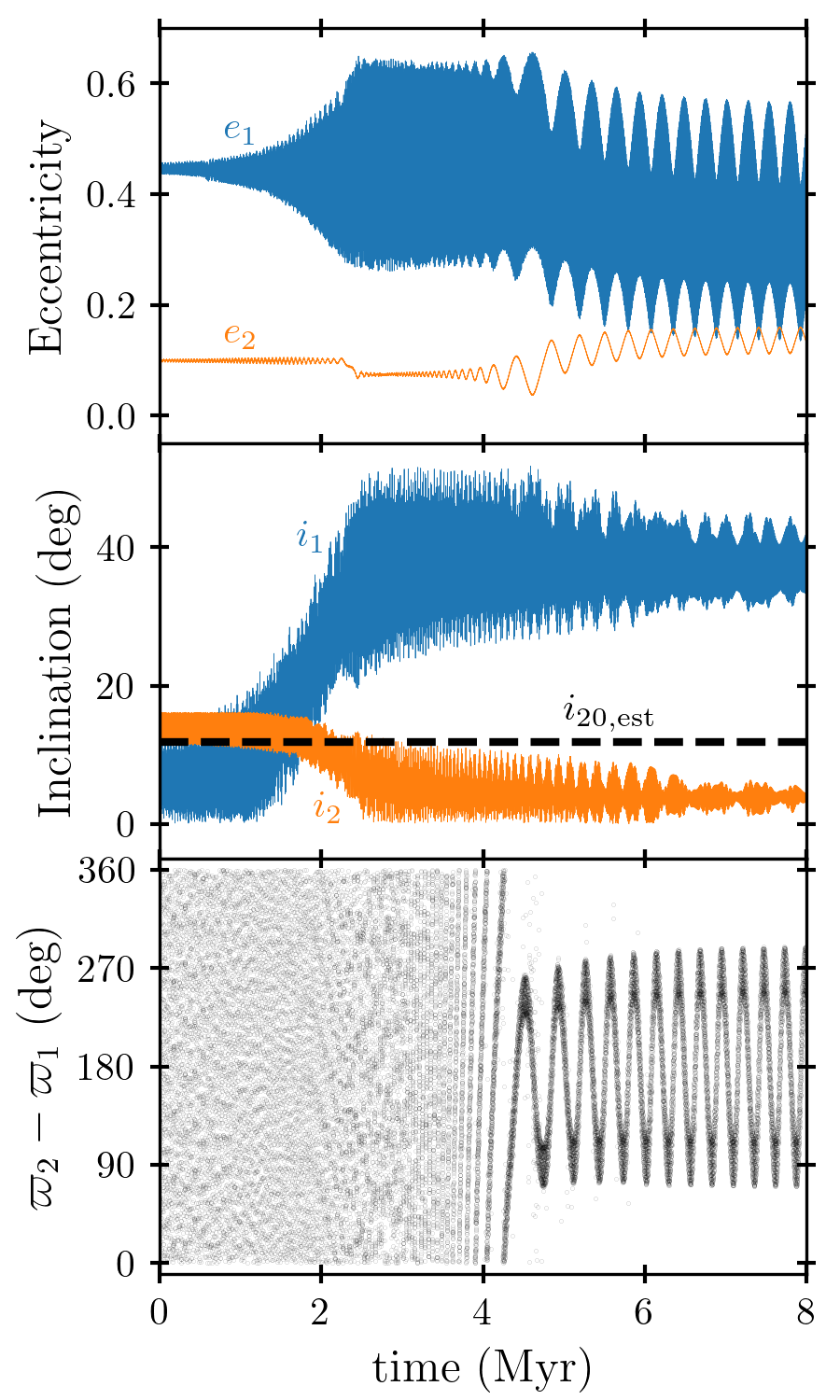}
   \caption{The mutual inclination between HD 147018b and c hypothesized by \protect\citetalias{DawsonChiang(2014)} to be $\sim$$39^\circ$ today could have arisen from a secular resonance crossing driven by a decaying outer disc. The evolution shown is the result of a backwards-integration starting at $t=8 \ \Myr$ with ``initial'' conditions from \protect\citetalias{DawsonChiang(2014)} --- \{$M_\star$, $m_1$, $m_2$, $a_1$, $a_2$, $e_{\rm 1f}$, $e_{\rm 2f}$, $i_{\rm 1f}$, $i_{\rm 2f}$, $\om_{\rm 1f}$, $\om_{\rm 2f}$, $\Om_{\rm 1f}$, $\Om_{\rm 2f}$\} = \{$0.92 \, \Msun$, $2.1 \ \Mjup$, $6.6 \, \Mjup$, $0.24 \, \au$, $1.9 \, \au$, $0.47$, $0.13$, $35.6^\circ$, $3.4^\circ$, $66^\circ$, $136.9^\circ$, $0^\circ$, $180^\circ$\} --- augmented with a disc of mass $m_{\rm d} = 0.5 \, \Msun \exp(-t/t_{\rm d})$ and having parameters \{$\rin$, $\rout$, $t_{\rm d}$\} = \{$3.3 \, \au$, $150 \, \au$, $1 \ \Myr$\}.  The system may have originated with a mutual inclination of $i_{\rm 20, est} \approx 12^\circ$ (eqn.~\ref{eq:i20_est}), crossed a secular resonance that amplified $i_1$ at the expense of $i_2$, and settled into apsidal libration with $\Dg \varpi = \varpi_2-\varpi_1$ lingering near $\pm$90$^\circ$ (\protect\citetalias{DawsonChiang(2014)}). However, this formation scenario does not explain the origin of the  eccentricities in HD 147018b and c, which do not change much from their large present-day values.
   } 
\label{fig:DawsonChiang14_form}
\end{figure}

\citet[][hereafter \citetalias{DawsonChiang(2014)}]{DawsonChiang(2014)} identified a subset of warm Jupiters whose sky-projected arguments of pericenter $\omega_{\rm sky,1}$ differed from those of their exterior giant-planet companions by $|\Dg \om_{\rm sky}| \equiv |\om_{\rm sky, 1} - \om_{\rm sky, 2}| \approx 90^\circ$. Although $\om_{\rm sky}$ is technically a sky-projected angle (between the orbit's eccentricity vector and the vector from the star to the orbit's ascending node on the sky plane; see, e.g., figure 1 of \citealt{Chiang+(2001)}), \citetalias{DawsonChiang(2014)} found that $|\Dg \om_{\rm sky}|$ is a good proxy for $|\Dg \pom| = |\pom_2 - \pom_1|$, the angle between the eccentricity vectors of two orbits. Thus $|\Dg \om_{\rm sky}| \approx 90^\circ$ points to planet pairs with near-orthogonal apsides. Such pairs were argued by \citetalias{DawsonChiang(2014)} to have mutual inclinations of $\sim$40$^\circ$, in contrast to apsidally-aligned ($|\Dg \om_{\rm sky}| \approx 0^\circ$) and anti-aligned ($|\Dg \om_{\rm sky}| \approx 180^\circ$) pairs argued to be more nearly co-planar \citep[e.g.][]{Chiang+(2001),Nagasawa+(2003),Petrovich+(2019)}.

Figure~\ref{fig:Peri_Data} updates figure 1 of \citetalias{DawsonChiang(2014)}, showing $|\Delta \omega_{\rm sky}|$ for systems known to have two (and only two) planets with measured radial velocities or transit timing variations, taken from the NASA Exoplanet Archive (77\%), the Exoplanet Encyclopedia (18\%), and the Exoplanet Orbit Database (5\%). The clustering of systems with $|\Delta \omega_{\rm sky}| \approx 90^\circ$ noted by \citetalias{DawsonChiang(2014)} is no longer apparent.\footnote{\citetalias{DawsonChiang(2014)} drew the data for their figure 1 from the Exoplanet Orbit Database (EOD) only; the systems they highlighted as apsidally orthogonal are shown in red in Fig.~\ref{fig:Peri_Data}, with updated parameters. When we also restrict our sample to the EOD, we see evidence for the same clustering of $|\Dg \omega_{\rm sky}|$ near 90$^\circ$ that they reported. It is only when we add the data from the NASA Exoplanet Archive and the Exoplanet Encyclopedia that the clustering goes away.}  The histogram of $|\Delta \omega_{\rm sky}|$ on the right panel shows a fairly smooth continuum, with a mild preference for apsidal alignment over anti-alignment by a factor of $\sim$2-3.

What mutual inclinations do these $|\Dg \omega_{\rm sky}|$'s imply? To investigate this question, we integrate the secular equations of motion for two eccentric planets, systematically varying the initial mutual inclination $i_{\rm mut,0}$ and examining its effect on the phase-mixed distributions of $\Dg \pom = |\pom_2 - \pom_1|$ (which we use as a proxy for $\Dg \omega_{\rm sky}$, following  \citetalias{DawsonChiang(2014)}). We use equations (17)-(20) of \cite{Liu+(2015)} which can accommodate large eccentricities and inclinations, and incorporate general relativistic precession for the inner planet \citep[e.g.][]{EggletonKiseleva-Eggleton(2001)}. Parameters/initial conditions are chosen to be representative of giant planet pairs with radial velocity data: \{$m_1$, $a_1$, $e_{10}$\} = \{$1 \, \Mjup$, $0.2 \, \au$, $0.3$\},  \{$m_2$, $a_2$, $e_{20}$\} = \{$3 \, \Mjup$, $1.0 \, \au$, $0.3$\},  $\Om_{20} - \Om_{10} = 180^\circ$, and  $\om_{10}$ and $\om_{20}$ drawn uniformly from 0 to $2\pi$. 

The distributions of $|\Delta \pom|$ after 1 Myr, parameterized by $i_{\rm mut,0}$, are shown in Figure \ref{fig:Hist_PlanetPeri}. We see that the $|\Delta \pom|$ distributions corresponding to nearly co-planar systems, having a factor-of-2 preference for alignment over anti-alignment when  $i_{\rm mut,0} \lesssim 20^\circ$, seem to fit the observed distribution of $|\Dg \omega_{\rm sky}|$ (as shown in Fig.~\ref{fig:Peri_Data}) best. We conclude that the statistical evidence presented by \citetalias{DawsonChiang(2014)} for a population of highly inclined, apsidally-orthogonal giant-planet pairs no longer exists.

Nonetheless, it is still possible that a given individual system observed today to be apsidally orthogonal has a $\sim$40$^\circ$ mutual inclination, following the dynamics described by \citetalias{DawsonChiang(2014)} whereby $|\Dg \omega_{\rm sky}|$ lingers near 90$^\circ$ as it oscillates about 180$^\circ$. We now ask whether the large mutual inclination presumed for such configurations may have originated from a secular resonance crossing driven by a decaying outer disc. We take as a case study one of the systems highlighted by \citetalias{DawsonChiang(2014)}, HD 147018. To decide whether a crossing may have occurred, we literally integrate the system {\it backwards} in time, starting the calculation at $t = 8$ Myr (the present day),  ending at $t = 0$, and prescribing the outer disc to {\it increase} up to a mass of $m_{\rm d0} = 0.5 \, {\rm M}_\odot$. The disc is otherwise modeled the same way as in section \ref{sec:InnerPlanet}, for assumed parameters $\rin = 3.3 \, \au$, $\rout = 150 \, \au$, and $t_{\rm d} = 1 \ \Myr$.  We have verified that the system is time-reversible by integrating forwards and backwards and achieving consistent results.
We use again \cite{Liu+(2015)}
to model the secular interaction of two planets and include general relativistic precession for the inner planet. The contributions to planet precession from the disc are given by:
\begin{align}
    \frac{\der {\bm j}_k}{\der t} \bigg|_{k\der} &= f_{k \der} {\bm j}_k \btimes \hld \\
    \frac{\der {\bm e}_k}{\der t} \bigg|_{k\der} &= - f_{k \der} \left( 2 \ve_k \btimes {\bm j}_k - \ve_k \btimes \hld \right) 
\end{align}
\citep[e.g.][]{PuLai(2018),Petrovich+(2019)}. Here $\hl$ is the unit-vector orbit normal (subscript d for disc, and $k$ for planet $k$),  $\ve$ is the eccentricity vector of magnitude $e$ pointing in the direction of pericentre, and ${\bm j} = \sqrt{1 - e^2} \hl$. The frequency $f_{k\der}$ is given by equation~\eqref{eq:omkd}, $f_{\der k} = (L_k/L_\der)f_{k\der}$, $L_k = m_k \sqrt{G M_\star a_k}$, and $L_\der = (2/3) m_\der \sqrt{G M_\star \rout}$. The disc evolves according to:
\begin{align}
    \frac{\der \hld}{\der t} = f_{\der 1} \hld \btimes {\bm j}_1 + f_{\der 2} \hld \btimes {\bm j}_2 \,.
    \end{align}

Figure~\ref{fig:DawsonChiang14_form} displays the results of the back-integration of HD 147018, assumed to have a present-day mutual inclination of $i_{\rm mut} = 39^\circ$. The system may indeed have once crossed a secular resonance which increased $i_{\rm mut}$ by a factor of $\sim$3. In this scenario, the inclination $i_2$ of the outer planet needs to have been $\sim$10--15$^\circ$ in the past; this value agrees with that estimated by inverting equation~\eqref{eq:i1_max_LL}:
\begin{equation}
    i_{\rm 20, est} \simeq 2 \sin^{-1} \left( \sqrt{ \frac{L_1}{L_2} } \sin \frac{1}{2} i_{\rm mut} \right) \approx 12^\circ
    \label{eq:i20_est}
\end{equation}
shown as a dashed line in Fig.~\ref{fig:DawsonChiang14_form} for $i_{\rm mut} = 39^\circ$. By contrast, the eccentricities do not change as much; $e_1$ in the past needs to have been about as large as its present-day value of 0.47. Thus our scenario of a decaying outer disc does not explain the origin of the large eccentricities in HD 147018.

\subsection{Stellar Spin}
\label{sec:Extra_SOMis}

We consider here how accounting for the host stellar spin and quadrupole moment changes the dynamics of secular resonance crossing. Mostly we find that it does not, for our parameter space.

\subsubsection{Star + planet interactions}
\label{sec:SOMis_StarPlanet}

For a planet to substantively tilt the spin axis of a star, at least two conditions need to be satisfied. First, the planet's orbital angular momentum $L_{\rm p} = m_{\rm p} \sqrt{G M_\star a_{\rm p}}$ should be larger than the stellar spin angular momentum $S = k_\star M_\star R_\star^2 \Omega_\star$. For our parameters,
\begin{equation}
    \frac{S}{L_{\rm p}} = 1.1 \left( \frac{1 \, \Mjup}{m_{\rm p}} \right) \left( \frac{1 \, \au}{a_{\rm p}} \right)^{1/2} \left( \frac{3 \, \der}{P_\star} \right) \left( \frac{M_\star}{1 \, \Msun} \right)^{1/2} \left( \frac{R_\star}{2 \, \Rsun} \right)^2 
    \label{eq:SoverLp}
\end{equation}
where $\Omega_\star = 2\pi/P_\star$ is the star's rotation frequency, $P_\star$ is the star's rotation period, and the angular momentum constant $k_\star \simeq 0.2$ for a fully convective body \citep[e.g.][]{Chandrasekhar(1939),Lai+(1993)}.
For $a_{\rm p} \lesssim 1 \, \au$, $S \gtrsim L_{\rm p}$ and the planet cannot control the star's tilt. For $a_{\rm p} \gtrsim 1 \, \au$, we run up against the second condition, that the 
precession frequency of the star driven by the planet 
\begin{equation}
    f_{\rm \star p} = 2.2 \times 10^{-5} \left( \frac{3 \, \der}{P_\star} \right) \left( \frac{m_{\rm p}}{1 \, \Mjup} \right) \left( \frac{1 \, \au}{a_{\rm p}} \right)^3 \frac{{\rm deg}}{\yr}
\end{equation}
be shorter than the precession frequency of the planet; otherwise the planet orbit normal would vary too rapidly to coherently tilt the stellar spin axis \citep[e.g.][]{Lai(2014),ZanazziLai(2018b)}. In evaluating $f_{\rm \star p}$ we have taken the star's second gravitational moment to be $J_2 = k_q \Omega_\star^2/(G M_\star/R_\star^3)$, with $k_q \simeq 0.09$ \citep[e.g.][]{Lai+(1993)}. For our parameters, typical values of $\dot \Om_{\rm p}$ (see Figs.~\ref{fig:Nbody_LL_comp} \&~\ref{fig:Nbody_pars} for $\dot \Om_1$ and $\dot \Om_2$) exceed $f_{\rm \star p}$. 

What about the converse torque exerted by the oblate star on the planet, conceivably important when $S > L_{\rm p}$? The planet's precession frequency forced by the star is:
\begin{align}
    f_{\rm p\star} = 2.4 \times 10^{-5} \left( \frac{3 \, \der}{P_\star} \right)^2 \left( \frac{1 \, \au}{a_{\rm p}} \right)^{7/2} \left( \frac{M_\star}{1 \, M_\odot} \right)^{1/2} \left( \frac{R_\star}{1 \, \Rsun} \right)^2 \frac{{\rm deg}}{{\rm yr}}.
\end{align}
For $a_{\rm p} \gtrsim 0.1 \, \au$, $f_{\rm p\star}$ is lower than precession frequencies driven by planet-planet interactions (see, e.g., $\dot \pom$, $\dot \Om$ in Figs.~\ref{fig:Nbody_LL_comp} \&~\ref{fig:Nbody_pars}). Thus the planets generally affect one another more than they are affected by the star.

\subsubsection{Star + inner disc interactions}
\label{sec:SOMis_StarDisk}

Our modeled inner disc typically has an angular momentum 
$L_1 = (2/3) m_1 \sqrt{G M_\star r_{\rm 1,out}}$ that is less than that of the star $S$. This fact, together with the inner disc extending down to small stellocentric radii, opens up the possibility that the star controls the precession rate of the inner disc. To the equations modeling the interaction between the inner disc, planet, and outer disc (eqs.~\ref{eq:dhl1dt}-\ref{eq:dhl2dt}), we add the contribution from the star to the evolution of the inner disc's orbit normal $\hl_1$:
\begin{align}
    \frac{\der \hl_1}{\der t} \bigg|_{\rm 1\star} &= f_{\rm 1\star} (\hl_1 \bcdot \hs)\hl_1 \btimes \hs 
    \label{eq:dhl1dt_star} 
    \end{align}
where the unit stellar spin vector $\hs$ evolves according to:
\begin{align}
    \frac{\der \hs}{\der t} \bigg|_{\rm \star 1} &= f_{\rm \star 1} (\hs \bcdot \hl_1)\hs \btimes \hl_1 
    \label{eq:dhsdt}
\end{align}
with precession frequencies
\begin{align}
    f_{\rm 1\star} &= \frac{1}{L_1} \int_{r_{\rm 1,in}}^{r_{\rm 1,out}} \frac{3 \pi G M_\star R_\star^2 J_2 \Sg_1}{r_1^2} \der r_1 
    \label{eq:f_1s} \\
    f_{\rm \star 1} &= \frac{L_1}{S} f_{\rm 1\star} \,. \label{eq:f_s1} 
\end{align}
The effects of the star on the planet (exterior to the inner disc) and outer disc are small and neglected.

\begin{figure}
    \centering
    \includegraphics[width=1.0\columnwidth]{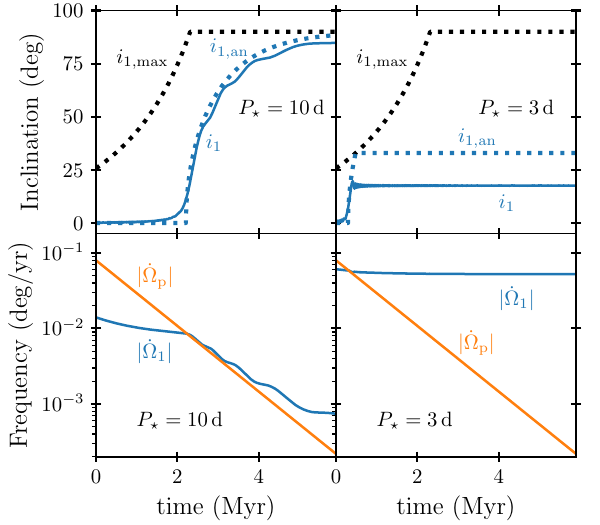}
    \caption{
    Similar to Fig.~\ref{fig:InnerDiskTilt}  showing the secular evolution of an inner disc + planet + outer disc, but now including the torque from the star's mass quadrupole for the stellar rotation periods $P_\star$ indicated. Other model parameters are \{$M_\star$, $m_{\rm 20}$, $\delta$, $r_{\rm 1,in}$, $r_{\rm 1,out}$, $r_{\rm 2,in}$, $r_{\rm 2,out}$, $m_{\rm p}$, $a_{\rm p}$\} = \{$2 \, \Msun$, $0.2 \, \Msun$, $8 \times 10^{-3}$, $0.03 \, \au$, $8 \, \au$, $20 \, \au$, $200 \, \au$, $5 \, \Mjup$, $16 \, \au$\}. For our parameters, including the stellar torque does not qualitatively change how the inner disc's inclination may be excited by an exterior planet. As the inner disc's precession rate $\dot \Om_1$ becomes increasingly dominated by the star ($f_{\rm 1\star} \gtrsim f_{\rm 1p}$), the secular resonance is crossed earlier (compare right column to left).
    }
    \label{fig:i1_star}
\end{figure}

Figure~\ref{fig:i1_star} shows the resultant evolution for two values of $P_\star$. We see that inner disc inclinations $i_1$ become excited by a secular resonance much as they did when stellar spin was ignored (Fig.~\ref{fig:InnerDiskTilt}). The star can increase substantially the nodal precession rate of the inner disc
\begin{equation}
    \dot \Om_1 \simeq -f_{\rm 1\star} \cos i_1 - f_{\rm 1p} \cos i_1 - f_{12} \cos i_1 \label{eq:dOm1_star}
\end{equation}
but the main consequence is just to cause the secular resonance to be crossed earlier (compare right and left columns of Fig.~\ref{fig:i1_star}). Our analytic estimate \eqref{eq:i1_an} for $i_1$ 
when $\dot \Om_1 = \dot \Om_{\rm p}$ becomes revised to 
\begin{align}
    \cos i_{\rm 1,an} \approx \frac{f_{\rm 2p}}{f_{\rm 1\star} + f_{\rm 1p} - f_{\rm p1} + f_{12}} \,.
    \label{eq:i1_an_star}
\end{align}
Further setting, as we did in section \ref{sec:InnerDisk}, $i_{\rm 1,an}(t) = i_{\rm 1,max}(t_{\rm c})$ where $t_{\rm c}$ is the time when $i_{\rm 1,an}$ first crosses $i_{\rm 1,max}$ (eqn.~\ref{eq:i1_max_d}),
predicts to within a factor of 2 the actual $i_1$.

\section{Summary and Discussion}
\label{sec:SummaryDiscussion}

We have investigated the secular dynamics of two giant planets encircled by a transition disc undergoing mass loss. We have also considered what happens when we replace the inner planet with a disc that is assumed to precess rigidly. We find that:
\begin{itemize}
    \item When the outer planet's angular momentum exceeds the inner planet's, the planets can cross nodal and apsidal secular resonances as the disc disperses. These crossings can magnify the orbital inclination of the inner planet relative to the outer planet, and the inner planet's eccentricity. The magnification factor is of order the square root of the ratio of the outer to inner planet's angular momentum 
    (eqs.~\ref{eq:e1_max_LL} and~\ref{eq:i1_max_LL}).     For typical massive two-planet systems, the magnification factor is on the order of a few  
    (Fig.~\ref{fig:Nbody_pars}). Thus, for example, generating a $\sim$40$^\circ$ mutual inclination may require a seed inclination of up to $\sim$10$^\circ$ (more on actual systems below). Whether or not a secular resonance is crossed depends in part on the outer disc mass; the crossings simulated in our work rely on outer disc masses initially comparable to the host stellar mass (similar to Class 0/I sources, e.g. \citealt{Jorgensen+(2009),Tobin+(2015),Segura-Cox+(2018),Andersen+(2019)}).
    \item Inner discs of the kind observed to reside within the cavities of transitional discs have such low mass 
    \citep{UbeiraGabellini+(2019),FrancisvanderMarel(2020),Portilla-Revelo+(2023)} and by extension low angular momentum that their inclinations relative to an exterior planet are more easily amplified by secular resonance. A seed mutual inclination of 
    $\sim$$3^\circ$ can grow
    to $\sim$$30^\circ$-$90^\circ$ (Fig.~\ref{fig:InnerDiskTilt}).
    This scenario may explain the tilted inner discs inferred to cast shadows on outer transition 
    discs (e.g.~\citealt{Benisty+(2022),Bohn+(2022)}; more on actual such systems below). 
    \item Planet pairs discovered through radial velocity measurements exhibit widely varying degrees of apsidal alignment, from aligned to anti-aligned and everything in between  (Fig.~\ref{fig:Peri_Data}). The apsidal distribution appears roughly consistent with such pairs predominantly residing on nearly co-planar orbits, with mutual inclinations $\lesssim 20^\circ$. Contrary to \cite{DawsonChiang(2014)}, we find no statistical evidence for a separate population of pairs on more highly inclined, apsidally-orthogonal orbits. Such orbits can still exist in principle and might describe 
    individual systems like HD 147018. This system's hypothesized large mutual inclination, but not its large eccentricities, may originate from a secular resonance crossing driven by a decaying outer disc.
    \item The scenarios we have explored are largely insensitive to the host star's spin, and vice versa. Thus when the inclination of an inner planet or inner disc is excited, the stellar spin axis does not follow suit; we expect large stellar obliquities.
\end{itemize}

\begin{figure}
    \centering
    \includegraphics[width=\columnwidth]{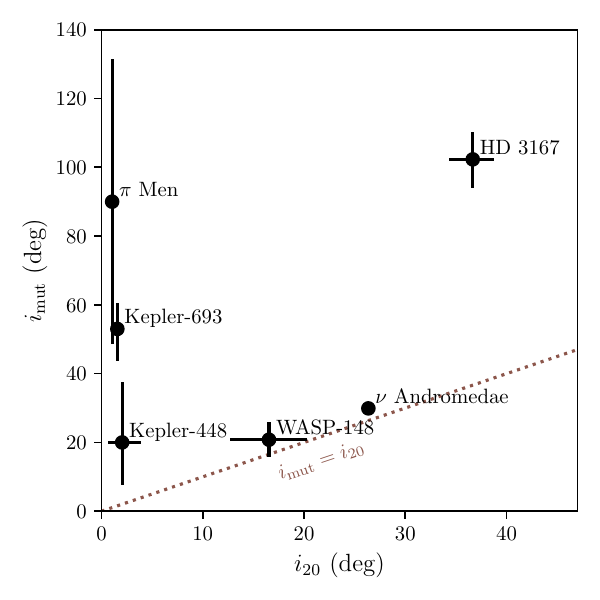}
    \caption{
    Seed inclination of outer planet $i_{20}$ required to generate the observed mutual inclination $i_{\rm mut}$ (eq.~\ref{eq:i20_est}).  Brown dotted line traces $i_{20} = i_{\rm mut}$. Systems like $\pi$ Men \protect\citep{XuanWyatt(2020)}, Kepler-448 and Kepler-693 \protect\citep{Masuda(2017)} have large angular momentum ratios $L_2/L_1$ and can therefore leverage small $i_{20}$ into large $i_{\rm mut}$ using a secular resonance. Other systems like WASP-148 \protect\citep{Almenara+(2022)} and $\nu$ Andromedae \protect\citep{McArthur+(2010)} have angular momentum ratios closer to unity and are therefore poor candidates for the secular amplification of mutual inclination. See text for discussion of the HD 3167 system \protect\citep{Bourrier+(2021)} which we hypothesize harbors a distant massive companion.  We do not plot KOI-984 \protect\citep{Sun+(2022)} because the mass and orbital period of the outer companion seem too uncertain; nor do we plot Kepler-108 \protect\citep{MillsFabrycky(2017)} because its angular momentum ratio $L_2/L_1<1$.
    }
    \label{fig:Inc_Data}
\end{figure}

We close our study by connecting to additional observed systems.  Given the angular momentum ratio of a two-planet system, equation~\eqref{eq:i20_est} estimates the seed inclination $i_{\rm 20}$ needed to produce a final mutual inclination $i_{\rm mut}$. Figure~\ref{fig:Inc_Data} plots $i_{\rm mut}$ vs. $i_{20}$ for systems with measured mutual inclinations. Kepler-448, Kepler-693, and $\pi$ Men are good candidates for disc-driven, secular excitation of their current inclinations, as they require relatively modest values of $i_{\rm 20} \lesssim 10^\circ$ (see also \citealt{Petrovich+(2020)}). In contrast, because angular momentum ratios are near unity in WASP-148 and $\nu$ Andromedae, a secular resonance can do little to magnify the systems' mutual inclinations ($i_{20} \simeq i_{\rm mut}$).  A more promising way to generate 
their inclinations may be planet-planet scattering \citep[e.g.][]{Chatterjee+(2008),JuricTremaine(2008),Barnes+(2011),Anderson+(2020)}. 

The HD 3167 system is particularly puzzling.  The planets HD 3167 b and c are transiting sub-Neptunes with orbital periods $P_{\rm orb} = 0.96 \, \der$ and $29.8 \, \der$, respectively. The inner member of the pair is nearly coplanar with the host star's equator, and the outer member is inclined by $i_{\rm mut} \approx 103^\circ$ \citep{Dalal+(2019),Bourrier+(2021)}.  Although a secular resonance can magnify the mutual inclination between these two planets by a factor of $\sim$4 (Fig.~\ref{fig:Inc_Data}), HD 3167 c does not have enough angular momentum to also tilt the star (eqn.~\ref{eq:SoverLp}).  A possible solution would be to have a massive distant companion (as yet unobserved) tilt the protoplanetary disc that formed HD 3167 c, leaving HD 3167 b unaffected because of its close proximity to the spinning star (section~\ref{sec:SOMis_StarDisk}).  In this scenario, we would also expect the other non-transiting sub-Neptunes in HD 3167 \citep{Bonomo+(2023)}, e ($P_{\rm orb} = 96.6 \, \der$) and perhaps also d ($P_{\rm orb} = 8.4\, \der$), to be approximately co-planar with c.

Other candidates for secular excitation of inclinations include Kepler-56, a system containing two planets of mass $0.07 \,\Mjup$ and $0.56 \,\Mjup$ and periods $10.5 \,\der$ and $21.4 \,\der$, whose orbits are co-planar with each other but inclined relative to the host star's equator by $\sim$$45^\circ$ \citep{Huber+(2013)}. A third body detected from radial velocity observations, with mass $m \sin i = 5.6 \, \Mjup$ and period $1000 \, \der$  \citep{Otor+(2016)}, may have tilted the disc that formed the inner two planets.  Another potential application of secular resonance is presented by the warm Jupiter TOI-1859b, hosted by a star with a projected obliquity of  $\lambda \approx 39^\circ$ \citep{Dong+(2023)}. An exterior companion larger than a few Jupiter masses (not yet observed), driven by a protoplanetary disc to precess at the right rate, could have misaligned TOI-1895b.

\begin{figure}
    \centering
    \includegraphics[width=1.0\columnwidth]{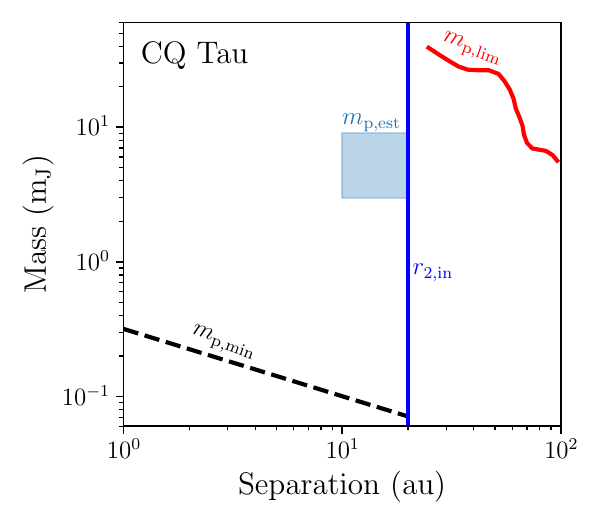}
    \caption{
    Constraints on the mass of a companion forming within the CQ Tau transition disc.  Red solid line plots the companion's maximum mass $m_{\rm p,lim}$ at a given location given coronograph non-detections \citep{Uyama+(2020),vanderMarel+(2021)}.  The blue solid line marks the estimated inner cavity edge $r_{\rm 2,in}$ \citep{UbeiraGabellini+(2019)}, while the light blue box encloses the estimated mass $m_{\rm p,est}$ and semi-major axis of a planet which can carve the disc cavity  \citep{UbeiraGabellini+(2019)}.  Black dashed line denotes the minimum companion mass ($m_{\rm p,min}$, eq.~\ref{eq:mp_min}) needed to misalign the inner disc by $i_1 = 44^\circ$ \citep{Bohn+(2022)}, for $m_2 \sim 2.9 \ {\rm m}_{\rm J}$, $r_{\rm 2,out} \sim 56 \ \au$, and $\delta \sim 10^{-2}$ \citep{UbeiraGabellini+(2019)}, and $i_{\rm p0} = 3^\circ$ and $r_{\rm 1,out} = 1 \ \au$ assumed arbitrarily.  Since $m_{\rm p,min} < m_{\rm p,est}$, the putative companion can misalign the inner disc and generate CQ Tau's shadows.
    }
    \label{fig:CQTau}
\end{figure}

CQ Tau hosts a transition disc with diametrically-opposed shadows \citep{Uyama+(2020)},  thought to be cast by an inner disc inclined by $44^\circ$ \citep{Bohn+(2022)}. Masses of both the inner and outer discs can be estimated from CO emission, and the mass and location of a planet interior to the disc cavity are constrained from coronagraphic observations and the need to shepherd the cavity edge
\citep{UbeiraGabellini+(2019),Uyama+(2020),vanderMarel+(2021)}. The constraints on the planet are plotted in Figure~\ref{fig:CQTau} (data colored blue and red), and are completely compatible with a planet that can tilt the inner disc by 44$^\circ$ (black dashed line).  
The planet mass needed to tilt the inner disc scales linearly with the inner disc mass (equation \ref{eq:mp_min}); the latter could be underestimated by nearly two orders of magnitude and still be consistent with our scenario of secular resonance crossing.

\section*{Acknowledgements}
Financial support was provided by NSF AST grant 2205500 and a 51 Pegasi b Heising-Simons Fellowship awarded to JJZ. We thank Rebekah Dawson and Cristobal Petrovich for discussions.

\section*{Data availability}

The data underlying this article will be shared on reasonable request to the corresponding author.

\appendix

\section{Sweeping Secular Resonance Model}
\label{app:SweepSecRes}

We construct a model for a sweeping secular resonance using Laplace-Lagrange theory.  We assume the nodal and apsidal precession periods are much shorter than the disc depletion time $t_{\rm d}$. Equations~\eqref{eq:dpom1dt_LL}-\eqref{eq:dpom2dt_LL} and~\eqref{eq:dOm1dt_LL}-\eqref{eq:dOm2dt_LL} describe how $\Dg \pom = \pom_2 - \pom_1$ and $\Dg \Om = \Om_2 - \Om_1$ evolve:
\begin{align}
    \frac{\der \Dg \pom}{\der t} = \Dg f - \left( g_{21} \frac{e_1}{e_2} - g_{12} \frac{e_2}{e_1} \right) \cos \Dg \pom 
    \label{eq:dDgpomdt_LL} \\
    \frac{\der \Dg \Om}{\der t} = -\Dg f + \left( f_{21} \frac{s_1}{s_2} - f_{12} \frac{s_2}{s_1} \right) \cos \Dg \Om 
    \label{eq:dDgOmdt_LL}
\end{align}
where
\begin{equation}
    \Dg f = f_{21} + f_{2\der} - f_{12} - f_{1\der}
\end{equation}
equals the difference between the two planets' non-oscillatory precession frequencies (eqs.~\ref{eq:dpom1dt_sec}-\ref{eq:dOm2dt_sec}).  For our decaying transition disc scenario, $\Dg f$ starts positive and decreases with time.

Equations~\eqref{eq:de1dt_LL}-\eqref{eq:de2dt_LL} and~\eqref{eq:ds1dt_LL}-\eqref{eq:ds2dt_LL} imply that eccentricities and inclinations are constant when apses and nodes are either aligned ($\Dg \pom, \Dg \Om = 0$) or anti-aligned ($\Dg \pom, \Dg \Om = \pi$).  Alignment ($+$) or anti-alignment ($-$) is enforced when 
\begin{align}
    \left. \frac{e_1}{e_2} \right|_\pm &= \frac{1}{2} \left[ \sqrt{ \left( \frac{\Dg f}{g_{21}} \right)^2 + 4 \frac{L_2}{L_1} } \pm \frac{\Dg f}{g_{21}} \right] 
    \label{eq:e1e2_lock} \\
    \left. \frac{s_1}{s_2} \right|_\pm &= \frac{1}{2} \left[ \sqrt{ \left( \frac{\Dg f}{f_{21}} \right)^2 + 4 \frac{L_2}{L_1} } \pm \frac{\Dg f}{f_{21}} \right] 
    \label{eq:s1s2_lock}
\end{align}
derived by setting  \eqref{eq:dDgpomdt_LL}-\eqref{eq:dDgOmdt_LL} to zero. Combining \eqref{eq:e1e2_lock}-\eqref{eq:s1s2_lock} with the conserved quantities \eqref{eq:e_const}-\eqref{eq:s_const} --- a.k.a.~the angular momentum deficit --- yields 
 $e_1$, $e_2$, $s_1$, and $s_2$ in apsidal/nodal lock. Notice $e_1/e_2|_\pm$ and $s_1/s_2|_\pm$ have the same value when $\Dg f = 0$:
\begin{equation}
    \left. \frac{e_1}{e_2} \right|_{\Dg f = 0} = \left. \frac{s_1}{s_2} \right|_{\Dg f = 0} = \sqrt{ \frac{L_2}{L_1} } \,. \label{eq:ratio_crit}
\end{equation}

\begin{figure}
    \centering  
    \includegraphics[width=0.8\linewidth]{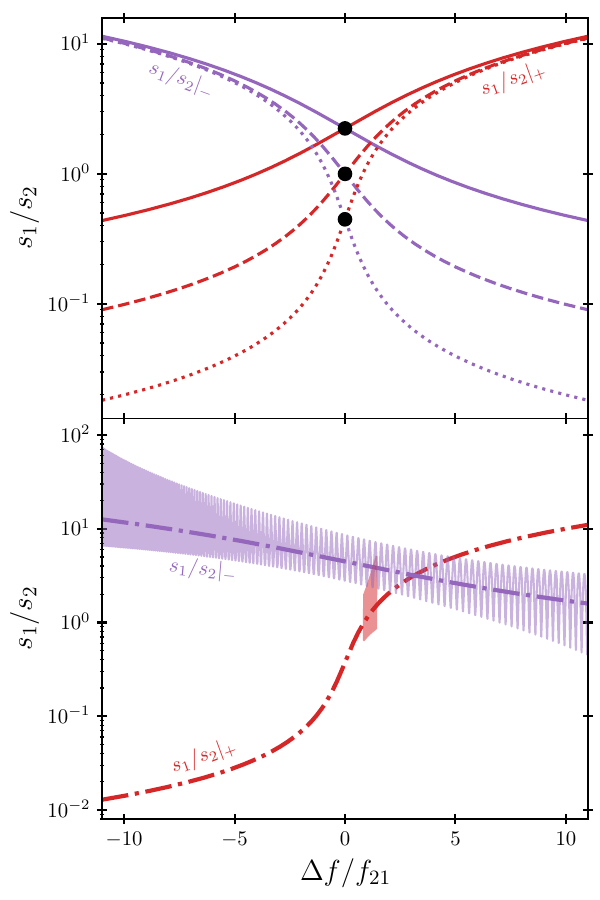}
   \caption{The inclination ratio $s_1/s_2$ vs.~$\Dg f/f_{21}$ (eq.~\protect\ref{eq:s1s2_lock}) for the nodally aligned ($+$, red) and anti-aligned ($-$, purple) modes.  \textit{Top panel}: 
 Aligned and anti-aligned solutions for $L_2/L_1 = 5, 1,$ and $0.2$ (solid, dashed, and dotted lines, respectively).  The solutions are symmetric about the critical value $s_1/s_2|_{\Dg f = 0} = \sqrt{L_2/L_1}$ (eq.~\ref{eq:ratio_crit}, solid circles). 
 \textit{Bottom panel}:  Comparing the full Laplace-Lagrange integrations in Fig.~\ref{fig:Nbody_LL_comp} (solid purple, $L_2/L_1 > 1$) and~Fig.~\ref{fig:Nbody_lowL} (solid red, $L_2/L_1 < 1$) to their respective anti-aligned (purple dot-dashed) and aligned (red dot-dashed) tracks. Time advances from right to left. The detuning frequency parameter $\Dg f$ does not cross zero when $L_2/L_1 < 1$, forestalling large changes to $s_1/s_2$.
   }
\label{fig:ResModel}
\end{figure}

From hereon we focus on how inclinations evolve; analogous statements apply for eccentricities. 
The top panel of Figure~\ref{fig:ResModel} shows  $s_1/s_2|_\pm$ vs.~$\Dg f$. In the beginning ($t=0$), $\Dg f > 0$.  Which of the aligned or anti-aligned solutions is relevant depends on $L_2/L_1$ and initial conditions (including the initial  $s_{10}/s_{20}$ and the nodes). As long as $\Dg f$ changes slowly, the system tends to move along a single $+$ or $-$ track (in reality oscillating about the track when nodal oscillations are non-zero).  
As $\Dg f$ decreases, $s_1/s_2|_+$ decreases and $s_1/s_2|_-$ increases, with $s_1/s_2$ changing most rapidly when $\Dg f$ crosses zero (secular resonance passage).
Because the angular momentum deficit is conserved, the relative $S_k = \frac{1}{2} L_k s_k^2$ values determine how the two orbits exchange inclination. The inner planet's inclination $s_1$ will amplify most (at the expense of the outer planet's $s_2$) upon secular resonance passage along an anti-aligned track, for $S_{10} < S_{20}$ initially and $S_{\rm 1f} > S_{\rm 2f}$ in the final state --- equivalently, when $s_{10}/s_{20} < \sqrt{L_2/L_1} < s_{\rm 1f}/s_{\rm 2f}$. The various examples given throughout this paper of mutual inclination excitation by secular resonance passage follow this anti-aligned, $L_2/L_1 > 1$ track --- a sample evolution taken from Fig.~\ref{fig:Nbody_LL_comp} is plotted in the bottom panel of Fig.~\ref{fig:ResModel}.

In principle, mutual inclination excitation is also possible by following an $L_2/L_1 < 1$, aligned (+) track to decrease $s_1$ and amplify $s_2$. But in practice, as explained at the end of section \ref{subsec:NN}, $\Delta f$ never crosses zero when $L_2/L_1 < 1$ in our decaying outer disc scenario, and therefore large changes to $s_1/s_2$ do not materialize. This is confirmed by the red trajectory, taken from Fig.~\ref{fig:Nbody_lowL}, plotted in the bottom panel of Fig.~\ref{fig:ResModel}.

\bibliographystyle{mnras}
\bibliography{main}

\end{document}